\documentclass[usegraphicx,usedcolumn,usenatbib]{mn2e}
\usepackage{color,epsfig,rotating,amssymb,longtable}
\usepackage{array,colortbl}
\usepackage[colorlinks=true,linkcolor=magenta,citecolor=blue,breaklinks=true]{hyperref}

\usepackage{ifpdf}

\ifpdf
\usepackage{epstopdf}
\else
\usepackage[dvips]{graphicx}
\usepackage{lscape}
\fi

\def\HII{H{\sc ii}}
\def\SII{S{\sc ii}}

\def\OIII{O{\sc iii}}

\def\Msol{M$_\odot$}

\usepackage{color}

\definecolor{dgreen}{rgb}{0,.5,.1} 
\definecolor{pink}{rgb}{.9,.2,.5}  
\definecolor{orange}{rgb}{.9,.4,0} 
\definecolor{darkred}{rgb}{.545,0.0,.0}

\begin{document}

\title
{The circumnuclear environment of the peculiar galaxy NGC\,3310}
\author[G. F. H\"agele et al.]
{Guillermo F. H\"agele$^{1,2}$\thanks{E-mail: guille.hagele@uam.es},
\'Angeles I. D\'{\i}az$^{1}$,
 M\'onica V. Cardaci$^{1,2,3}$\thanks{PhD fellow of Ministerio de Educaci\'on y
    Ciencia, Spain}, 
   \newauthor Elena Terlevich$^{4}$\thanks{Research
  Affiliate at IoA} and Roberto Terlevich$^{4}$\footnotemark[3]
\\
$^{1}$ Departamento de F\'{\i}sica Te\'orica, C-XI, Universidad Aut\'onoma de
Madrid, 28049 Madrid, Spain\\ 
$^{2}$ Facultad de Cs Astron\'omicas y Geof\'isicas, Universidad Nacional de La
Plata, Paseo del Bosque s/n, 1900 La Plata, Argentina \\ 
$^{3}$ XMM Science Operations Centre, European Space Astronomy Centre of ESA,
P.O. Box 50727, 28080 Madrid, Spain\\ 
$^{4}$ INAOE, Tonantzintla, Apdo. Postal 51, 72000 Puebla, M\'exico\\ }

\maketitle

\begin{abstract}


Gas and star velocity dispersions have been derived for eight circumnuclear 
star-forming regions (CNSFRs) and the nucleus of the spiral galaxy
NGC\,3310 using high resolution spectroscopy in the blue and far red. Stellar
velocity dispersions have been obtained from the Ca{\sc ii} triplet in the
near-IR, using cross-correlation techniques,
while gas velocity dispersions have been measured by Gaussian fits to the
H$\beta$\,$\lambda$\,4861\,\AA\ and [\OIII]\,$\lambda$\,5007\,\AA\ emission
lines.

The CNSFRs stellar velocity dispersions range from 31 to
73\,km\,s$^{-1}$. These values, together with the sizes measured on archival
HST  
images, yield upper limits to the dynamical masses for the individual star
clusters between 1.8 and 7.1\,$\times$\,10$^6$\,M$_\odot$, 
for the whole CNSFR between 2\,$\times$\,10$^7$ and
1.4\,$\times$\,10$^8$\,M$_\odot$, and 5.3\,$\times$\,10$^7$\,M$_\odot$ for
the nucleus inside the inner 14.2\,pc.
The masses of the ionizing stellar population responsible for the \HII\ region
gaseous emission have been derived from their published H$\alpha$ luminosities
and are found to be between 8.7\,$\times$\,10$^5$ and
2.1\,$\times$\,10$^6$\,M$_\odot$ for the 
star-forming regions, and 2.1\,$\times$\,10$^5$\,M$_\odot$ for the galaxy
nucleus; they  therefore constitute between 1 and 7 per cent of the total dynamical
mass. 

The ionized gas kinematics is complex; two different kinematical components
seem to be present as evidenced by different line widths and Doppler shifts.

\end{abstract}


\section{Introduction}

The gas flows in disc of spiral galaxies can be strongly perturbed by the
presence of bars, although the total disc star formation rates (SFR) do not
appear to be 
significantly affected by them \citep{1998ARA&A..36..189K}. These perturbations
of the gas flow trigger nuclear star formation in the bulges of some barred
spiral galaxies. 
External environmental influences can have strong effects on the SFR, among
them, the most important by far, 
is tidal interactions.
The young extragalactic star clusters belonging to these systems
have been the aim of different studies during the last decades
\citep[e.g.\
][]{1991MNRAS.253..245D,1993AJ....106.1354W,2005A&A...443...41M,2005A&A...431..905B,2006A&A...448..881B,2008A&A...489.1091M}. 
The enhancement of the 
SFR is highly variable depending on the star formation conditions, the degree
of enhancement ranging from zero in gas-poor galaxies to around 
10-100 times in extreme cases \citep{1998ARA&A..36..189K}. Much larger
enhancements are often seen in the circumnuclear regions of strongly
interacting and merging systems
\citep{1998ARA&A..36..189K,1998ApJ...498..541K}. 

Yet, the bulges of some nearby, non-interacting, spiral galaxies show intense
star-forming regions located in a  roughly annular pattern around their
nuclei. In the middle of last century, \cite{1958PASP...70..364M} 
classified a sample of galaxies using as the principal classification 
criterion the degree of central concentration of light of each
galaxy. An apparent fairly common phenomenon in some types of galaxies was
pointed out by Morgan: their nuclear regions can consist of
an extremely bright, small nucleus superposed on a considerably fainter
background, or it may be made up of multiple ``hot-spots''. Almost a decade
later, 
\cite{1965PASP...77..287S} suggested a relationship between the existence
of a bar and the presence of abnormal features in their nuclei for a survey of
bright southern galaxies. These authors extended the survey to the whole sky
\citep{1967PASP...79..152S}, and found that
$\sim$\,14\,\% of these galaxies presented peculiar nuclei. 
The distinctive nature (with respect
to the more extended star formation in discs) of the luminous nuclear
star-forming regions was fully revealed with the opening of the mid- and
far-infrared (IR) spectral ranges 
\citep[see e.g.\
][]{1972ApJ...176L..95R,1973ApJ...182L..89H,1978ApJ...220L..37R,1980ApJ...235..392T}. 

In general, CNSFRs and giant \HII\ regions in the discs of galaxies are very
much alike, although  
the former look more compact and show higher peak surface brightness
\citep{1989AJ.....97.1022K} than the latter. 
CNSFRs, with sizes going from a few tens to a few hundreds of parsecs 
\citep[e.g.][]{2000MNRAS.311..120D} seem to be made of several \HII\ regions
ionized by luminous compact stellar clusters whose sizes, as measured from
high spatial resolution {\it Hubble Space Telescope} (HST) images, are seen to
be of only a few parsecs. 
Their large H$\alpha$ luminosities, typically higher than
10$^{39}$\,erg\,s$^{-1}$, point to relatively
massive star clusters as their 
ionization source.
Although these \HII\ regions are very luminous (M$_v$ between -12 and -17) not
much is known about their  
kinematics or dynamics for both the ionized gas and the stars. It could be
said that the worst known parameter of these ionizing clusters is their mass. 
As derived with the use of population synthesis models their masses suggest
that these  clusters are gravitationally bound and that they might evolve into
globular cluster configurations \citep{1996AJ....111.2248M}. {\bf Further, deeper analysis as to whether or not such cluster would survive the hostile environment of the circumnuclear regions is extremely interesting but lies  outside the scope of the present work.
}
Classically it is assumed that the system is virialized 
hence
the total mass inside a radius can be determined by applying the virial theorem to the observed 
velocity dispersion of the stars ($\sigma_{\ast}$). 
As pointed out by several authors \citep[e.g.\ ][]{1996ApJ...466L..83H}, at 
near-IR
wavelengths (\,8500\,\AA) the contamination due to nebular lines is much
smaller  and since red supergiant stars, if present, dominate the
light where the Ca{\sc ii} $\lambda\lambda$\,8498, 8542, 8662\,\AA\ triplet
(CaT) lines are found, these should be easily observable allowing the
determination of $\sigma_{\ast}$
\citep{1990MNRAS.242P..48T,1994A&A...288..396P}. 

The equivalent width of the emission lines are lower than those shown by the
disc \HII\ regions \citep[see e.g.\
][]{1989AJ.....97.1022K,1997AJ....113..975B,1999ApJ...510..104B}.  
Combining GEMINI data and a grid of photo-ionization models
\cite{2008A&A...482...59D} conclude that the contamination of the continua
of CNSFRs by underlying contributions from both old bulge stars and stars
formed in the ring in previous episodes of star formation (10-20\,Myr) yield
the observed low equivalent widths.

This is the third paper of a series to study the peculiar conditions of
star formation in circumnuclear regions of early type spiral galaxies, 
in particular the kinematics of the connected stars and gas.
In this paper we present high-resolution  far-red spectra
and  stellar velocity dispersion
measurements ($\sigma_{\ast}$) along the line of sight for eight
CNSFRs and the nucleus of the spiral galaxy NGC\,3310. 

NGC\,3310 (UGC\,5786, Arp217) is a starburst galaxy
classified as an SAB(r)bc by 
\cite{1991trcb.book.....D}, with an inclination of the galactic disc of
about i\,$\sim$\,40 \citep{2000MNRAS.312....2S}. 
Its coordinates are $\alpha_{\rm 2000}$=10$^h$\,38$^m$\,45\fs9, 
$\delta_{\rm 2000}$=+53$^{\circ}$\,30\arcmin\,12\arcsec
\citep{1991trcb.book.....D}. These authors derived a distance to the galaxy
equal to 15\,Mpc, giving a linear scale of $\sim$\,73\,pc\,arcsec$^{-1}$.
This galaxy is a good example of an overall low 
metallicity galaxy, with a high rate of star formation and very blue
colours. This galaxy has a ring of star forming regions whose 
diameter ranges from 8\,\arcsec to 12\,\arcsec and shows two tightly
wound spiral arms \citep{2002AJ....123.1381E,1976A&A....48..373V} filled with
giant \HII\ regions. These circumnuclear regions present low metal abundance
\citep[0.2-0.4\,Z$_\odot$][]{1993MNRAS.260..177P}, in contrast to what is
generally found in this type of objects, which show high metallicities
\citep{2007MNRAS.382..251D}. In fact, in most cases, the [O{\sc
iii}]\,$\lambda$\,5007\,\AA\ forbidden line can barely be seen 
\citep[see e.g.\ ][hereafter Paper I and Paper II,
  respectively]{2007MNRAS.378..163H,2009MNRAS.396.2295H}. 

The ages indicated by the colours and magnitudes of the
star formation regions are lower than 10\,Myr \citep{2002AJ....123.1381E}. From
near-IR J and K photometry these authors derived an average age of
$\sim$10$^7$\,yr for the large scale ``hot-spots'' (star forming
complexes). From the observed CaT line 
in the Jumbo \HII\ region \cite{1990MNRAS.242P..48T} derived an age around
5 to 6\,Myr. \cite{2002AJ....123.1381E}, comparing their data with
Starburst99 models \citep{1999ApJS..123....3L}, estimated masses of the large
``hot-spots'' ranging from 10$^4$ to several times 10$^5$\,M$_\odot$. They
found 
17 candidate super star clusters (SSCs) with absolute magnitudes between
M$_B$\,=\,-11 and -15\,mag, and with colours similar to those measured for SSCs
in other galaxies 
\citep[see for example][]{1995AJ....110.1009B,2001ApJ...556..801L}.

We have measured
the ionized gas velocity dispersions ($\sigma_{g}$) from high-resolution blue
spectra using Balmer H$\beta$ and [O{\sc iii}] emission
lines. The comparison between $\sigma_{\ast}$ and $\sigma_{g}$  {\bf on an ample sample of objects} might throw
some light on the yet unsolved issue about the validity of the gravitational
hypothesis for the origin of the supersonic motions observed in the ionized
gas in Giant \HII\ regions \citep*{1999MNRAS.302..677M}. 
In Section 2 we describe the observations and data reduction. We present the 
results in Section 3, the dynamical mass derivation in Section 4 and the
ionizing star cluster properties in Section 5. We discuss all our results in
Section 6. Finally, the summary and conclusions are given in Section 7.

\section{Observations and data reduction}
\label{Obs}


\begin{table*}
\centering
\caption[]{Journal of Observations}
\begin{tabular} {l c c c c c c c c}
\hline
 Date & {\bf Slit} & Spectral range &       Disp.          & R$^a_{\rm{FWHM}}$ & Spatial res.            & PA   & Exposure Time & {seeing$_{\rm{FWHM}}$} \\
         &     &   (\AA)          & (\AA\,px$^{-1}$) &  (\AA)   & (\arcsec\,px$^{-1}$) &  ($ ^{o} $) & (sec) & (\arcsec)   \\
\hline
2000 February 4 & {\bf S1} & 4779-5199  &  0.21  &  12500  &  0.38  &    52    &  3\,$\times$\,1200 & 1.2\\
2000 February 4 &    & 8363-8763  &  0.39  &  12200  &  0.36  &    52    &  3\,$\times$\,1200 & \\
2000 February 5 & {\bf S2} & 4779-5199  &  0.21  &  12500  &  0.38  &   100    &  4\,$\times$\,1200 & 1.6\\
2000 February 5 &    & 8363-8763  &  0.39  &  12200  &  0.36  &   100    &  4\,$\times$\,1200 &\\
\hline
\multicolumn{7}{l}{$^a$R$_{\rm{FWHM}}$\,=\,$\lambda$/$\Delta\lambda_{\rm{FWHM}}$}
\end{tabular}
\label{journal}
\end{table*}


The data were acquired in February 2000 using the two arms of the Intermediate
dispersion Spectrograph and Imaging System (ISIS) 
attached to the 4.2-m William Herschel Telescope (WHT) of the
Isaac Newton Group (ING) at the Roque de los Muchachos Observatory on the
Spanish island of La Palma. The CCD
detectors EEV12 and TEK4 were used for the blue and red arms with a factor of
2 binning in both the ``x" and ``y" directions with resultant spatial
resolutions of 0.38 and 0.36 \,arcsec\,pixel$^{-1}$ for the blue and red
configurations respectively. The H2400B and R1200R gratings were used to cover
the wavelength ranges from 4779 to 5199\,\AA\ ($\lambda_c$\,=\,4989\,\AA) in
the blue and  from 8363 to 8763\,\AA\ ($\lambda_c$\,=\,8563\,\AA) in the red
with resultant spectral dispersions  of 0.21 and 0.39 \AA\ per pixel
respectively, providing a comparable velocity resolution of about
13\,km\,s$^{-1}$. A slit width of 1\,arcsec was used which,
combined with the spectral dispersions, yielded spectral resolutions of about
0.4 and 0.7\,\AA\ FWHM in the blue and the red, respectively, measured on the
sky lines. Table \ref{journal} summarizes the instrumental configuration and
observation details. 

\label{Data reduction}
The data were processed and analyzed using {\sc iraf}\footnote{{\sc iraf}: the
  Image 
Reduction and Analysis Facility is distributed by the National Optical
Astronomy Observatories, which is operated by the Association of Universities
for Research in Astronomy, Inc. (AURA) under cooperative agreement with the
National Science Foundation (NSF).} routines in the usual manner.
Further details concerning each step can be found in Paper I.
With the purpose of measuring radial velocities and velocity dispersions,
spectra of 11 template velocity stars were  acquired to provide good stellar
reference frames in the same system as the galaxy spectra for the kinematic
analysis in the far-red. They correspond to late-type giant and supergiant 
stars which have strong CaT features \citep[see][]{1989MNRAS.239..325D}. The
spectral types, luminosity classes and dates of observation of the stellar  
reference frames used as templates are listed in Table 2 of Paper I.

\label{Results}
\section{Results}

Two different slit positions (S1 and S2) were chosen in order to observe 8
CNSFRs and the nucleus of the galaxy. One of them the conspicuous
Jumbo region, labelled J, is the same region labelled  R19 by \cite{2000MNRAS.311..120D} 
and region A of \cite{1993MNRAS.260..177P}. 
\cite{1981A&A....96..271B} dubbed it Jumbo, given its extreme luminosity
\citep[100 times more luminous in IR than 30 Dor in the Large Magellanic
  Cloud, ][]{1984ApJ...284..557T}.

\begin{figure*}
    \centering
    \includegraphics[width=0.48\textwidth]{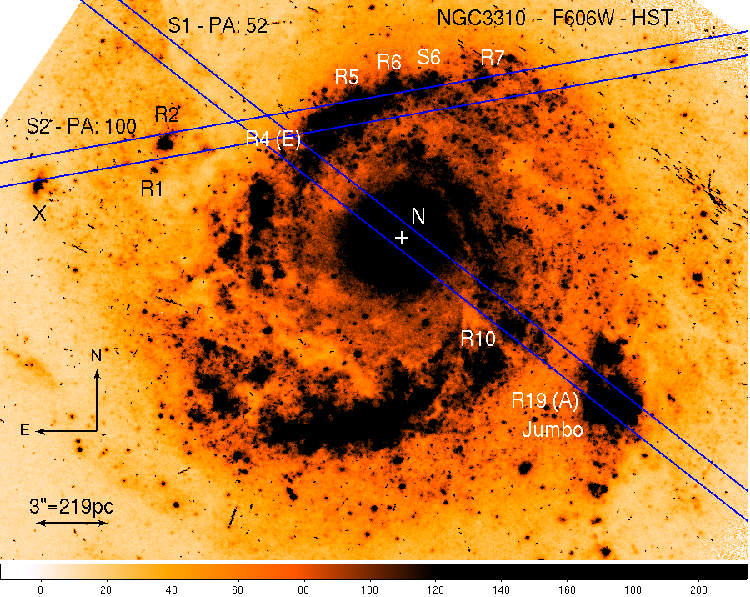}
\hspace{0.2cm}
\includegraphics[width=0.48\textwidth]{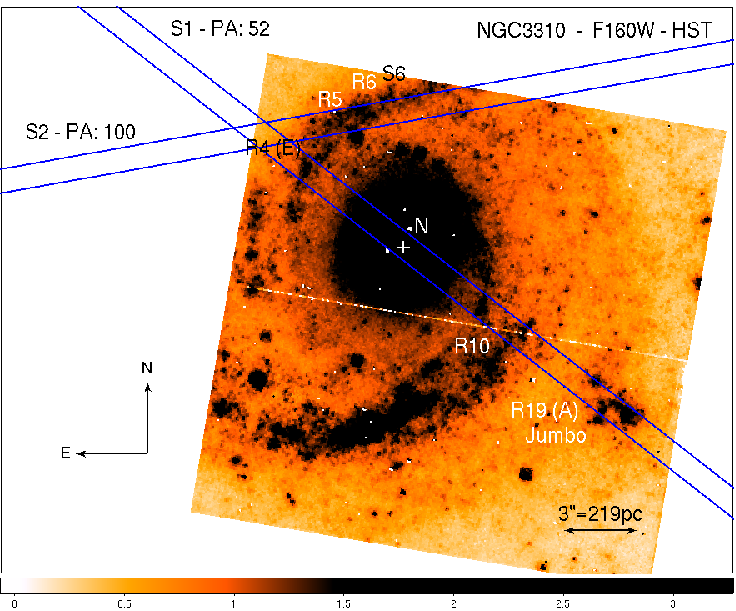}
\caption[]{Left: F606W (wide V) image centred on NGC\,3310 obtained with the
  WFPC2 camera (PC1) of the HST. Right: HST-NICMOS (NIC2) image obtained
  through the F160W filter. For both images the orientation is north up, east
  to the left. The location and P.A. of the WHT-ISIS slit positions, together
  with identifications of the CNSFRs extracted, are marked.}
\label{hst-slits}
\end{figure*}

\begin{figure}
\centering
\includegraphics[width=0.48\textwidth]{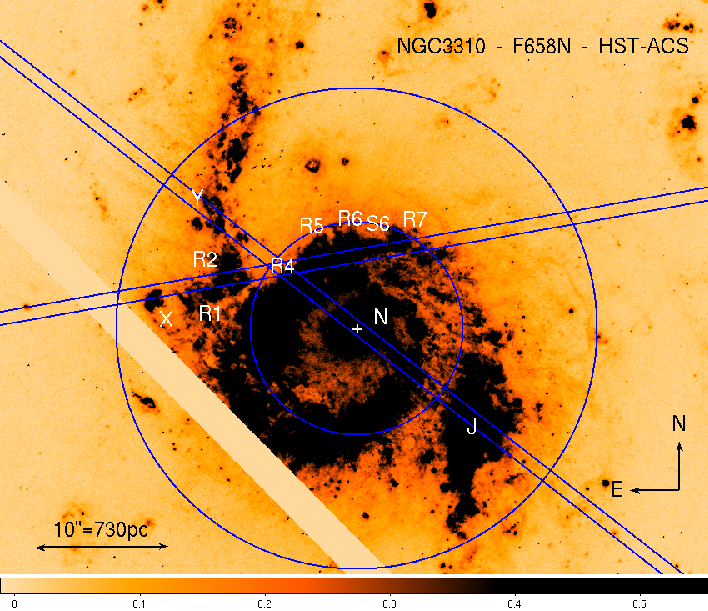}
\caption[]{F658N (narrow-band [N{\sc ii}] filter, at the redshift of
  NGC\,3310, z\,=\,0.0033 (\citealt{1998AJ....115...62H}), equivalent to the
  H$\alpha$ narrow band filter) image 
  centred on the galaxy obtained with the ACS camera of the HST. The
  orientation is north up, east to the left. The location and P.A. of the
  WHT-ISIS slit positions, together with identifications of the CNSFRs
  extracted, are marked. The radii of the circles, centred at the position of
  the nucleus, are 8 and 18\,\arcsec.}
\label{hst-slits-3310-ACS}
\end{figure}

Fig.\ \ref{hst-slits} shows the selected slits, superimposed on
photometrically calibrated optical and IR  images of the circumnuclear region
of this galaxy acquired with the Wide Field and Planetary Camera 2 (WFPC2;
PC1) and the Near-Infrared Camera and Multi-Object Spectrometer (NICMOS)
Camera 2 (NIC2) on board the HST. These images have been downloaded from the
Multimission 
Archive at STScI (MAST)\footnote{http://archive.stsci.edu/hst/wfpc2}. The
optical image was obtained through the F606W (wide V) filter, and the near-IR
one, through the F160W (H). The IR image does not cover the whole
circumnuclear ring, and it does not include the regions R1, R2, R7 and X. 
The CNSFRs have been labelled following the same
nomenclature as in \cite{2000MNRAS.311..120D}, with the nomenclature given by
\cite{1993MNRAS.260..177P} for the regions in common [E for R4 and A for R19
(the main knot of the Jumbo region)] within parentheses. In both  studies
the regions observed are identified on the H$\alpha$ maps. We have also
downloaded the F658N narrow band image (equivalent to 
H$\alpha$ filter at the redshift of NGC\,3310) taken with the HST Advanced Camera for
Surveys (ACS), shown in  Fig.\ \ref{hst-slits-3310-ACS}.
The plus symbols in both panels of Fig.\ \ref{hst-slits} and in Fig.\
\ref{hst-slits-3310-ACS} represent the position of the nucleus as given by
\cite{1999PASP..111..438F}.

\begin{figure*}
\centering
\includegraphics[width=.41\textwidth,angle=0]{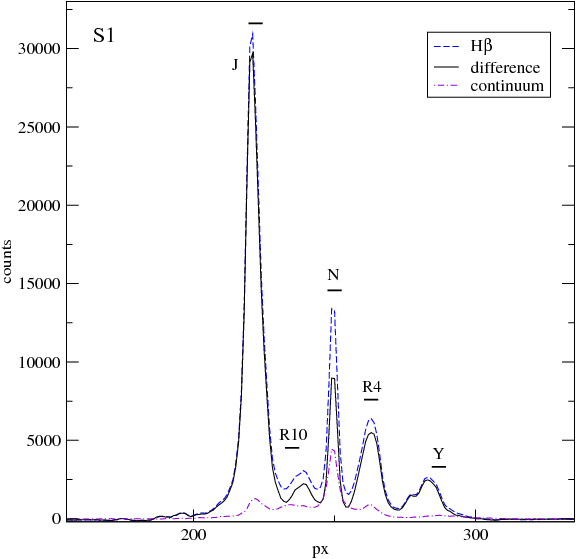}
\hspace{0.2cm}
\includegraphics[width=.41\textwidth,angle=0]{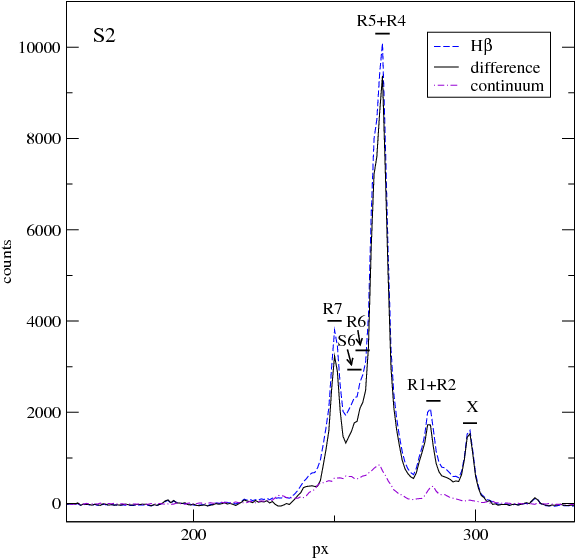}\\
\vspace{0.3cm}
\includegraphics[width=.41\textwidth,angle=0]{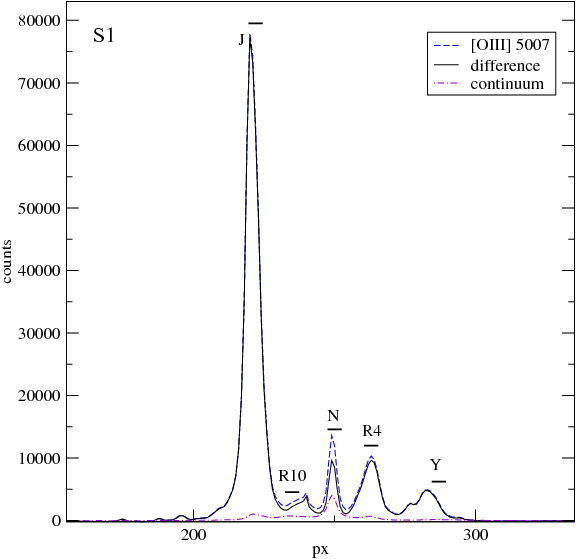}
\hspace{0.2cm}
\includegraphics[width=.41\textwidth,angle=0]{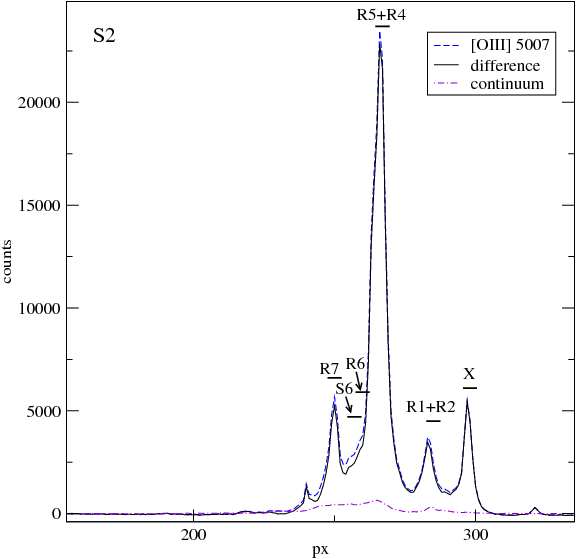}\\
\vspace{0.3cm}
\includegraphics[width=.41\textwidth,angle=0]{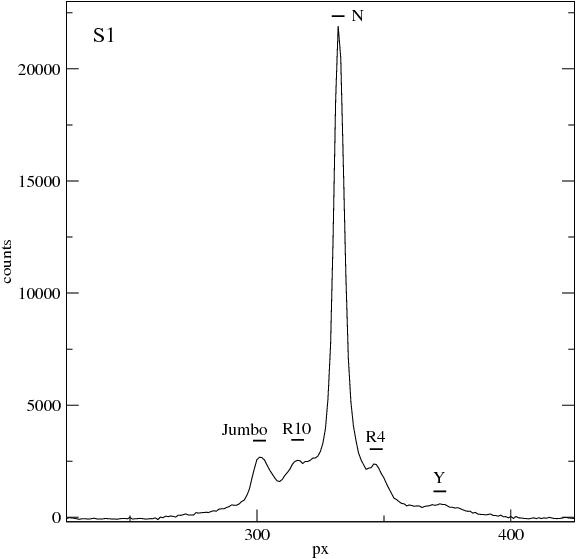}
\hspace{0.2cm}
\includegraphics[width=.41\textwidth,angle=0]{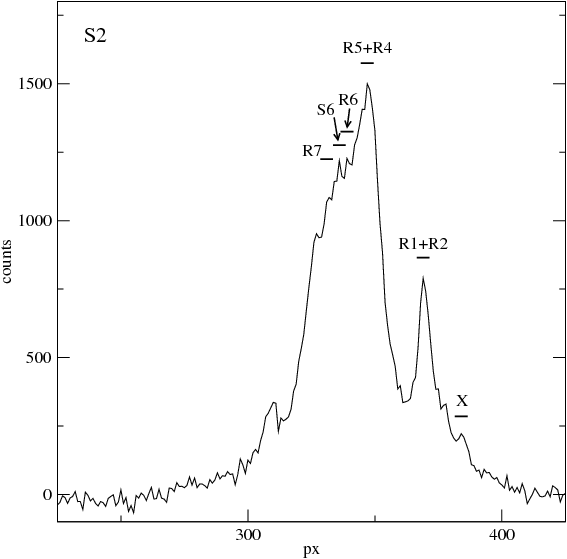}\\
\caption{Spatial profiles of H$\beta$, [O{\sc iii}]\,$\lambda$\,5007\AA\ and
  the far red (upper, middle and lower panels respectively) for each
  slit. For the emission lines, the profiles correspond to 
  line+continuum (dashed line), continuum (dashed-dotted line) and the 
  difference between them (solid line), representing the pure emission from
  H$\beta$ and [O{\sc iii}] respectively. For the far red profiles, the solid
  lines represent the continuum. Pixel number increases to the
  North. Horizontal small lines show the location of the CNSFRs and nuclear
  apertures.}
\label{profiles3310}
\end{figure*}

\begin{figure*}
\centering
\includegraphics[width=.41\textwidth,angle=0]{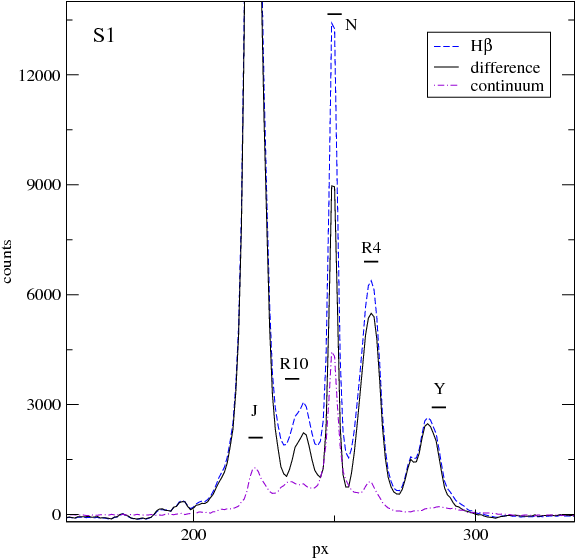}
\hspace{0.2cm}
\includegraphics[width=.41\textwidth,angle=0]{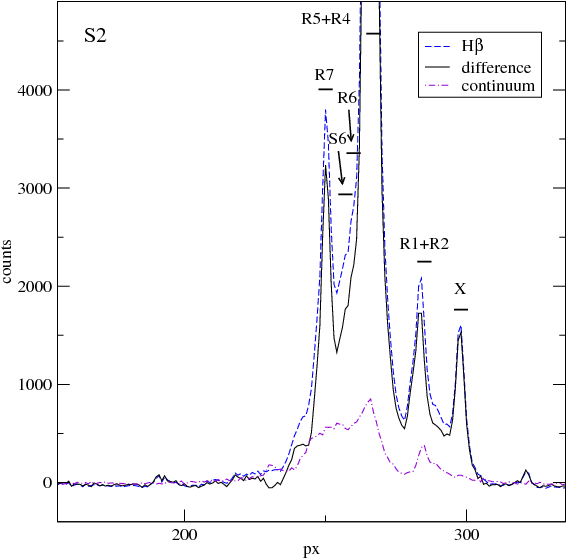}\\
\vspace{0.3cm}
\includegraphics[width=.41\textwidth,angle=0]{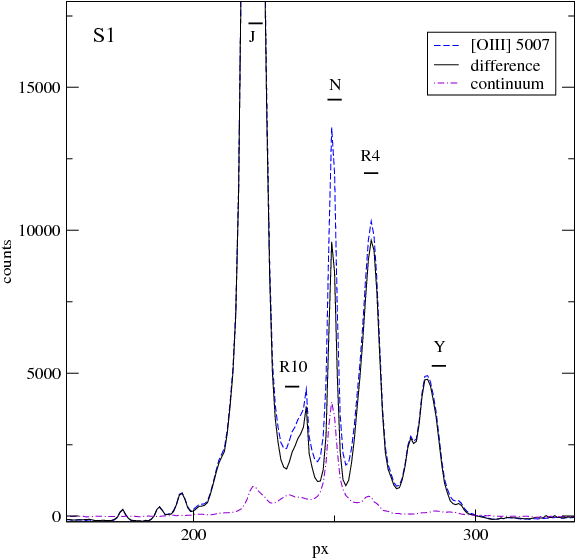}
\hspace{0.2cm}
\includegraphics[width=.41\textwidth,angle=0]{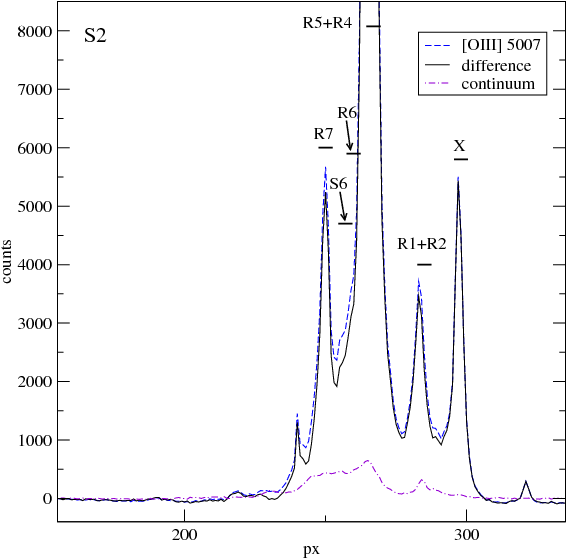}\\
\vspace{0.3cm}
\includegraphics[width=.41\textwidth,angle=0]{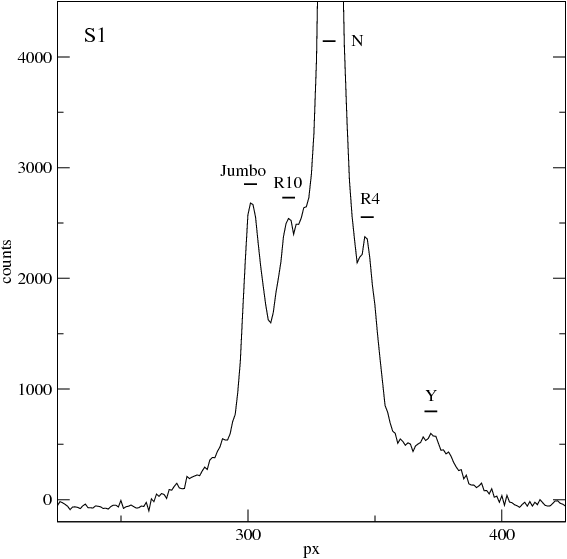}
\hspace{7.4cm}
\caption{Enlargement of the spatial profiles presented in
  Fig.\ \ref{profiles3310}, except in the far red range of S2.} 
\label{profiles3310-2}
\end{figure*}

Fig.\ \ref{profiles3310} shows the
spatial profiles in the H$\beta$ and  [O{\sc iii}]\,5007\,\AA\ emission lines
(upper and middle panels) and the far-red continuum (lower panel) along each
slit position. Due to
the presence of intense regions (J in the blue and the galaxy nucleus
in the red range in the case of S1, and R5+R4 in the blue range for S2) the
profile details are very 
difficult to appreciate, therefore we show some enlargements of these profiles
in Fig.\ \ref{profiles3310-2}. In all cases, the emission line
profiles have been generated by collapsing 11 pixels of the spectra in the
direction of the resolution at the central position of the lines in the rest
frame, $\lambda$\,4861 and $\lambda$\,5007\AA\ respectively, and are plotted
as dashed lines. Continuum profiles were generated by collapsing 11 resolution
pixels centred at 11\,\AA\ to the blue of each emission line and are
plotted  
as dash-dotted lines. The difference between the two, shown by a solid line, 
corresponds to the pure emission. The far-red
continuum has been generated by collapsing 11 pixels centered at
$\lambda$\,8620\AA.

The regions of the frames to be extracted into one-dimensional spectra 
corresponding to each of the identified CNSFRs, were selected on the continuum
emission profiles both in the blue and in the 
red. These regions are marked by horizontal lines and labelled in the
corresponding Figures. In the H$\beta$ profiles we find two almost pure
emission 
knots, one for each slit position, labelled Y and X by us (see Fig.\
\ref{hst-slits-3310-ACS}), respectively. The former of these regions seems
to be located at the tip of one vertical arm formed by pure emission regions,
since it can be easily appreciable in the H$\alpha$ image from the ACS (Fig.\
\ref{hst-slits-3310-ACS}) but it 
is almost invisible in the WFPC2-PC1 V-band image (see Fig.\
\ref{hst-slits}).

The spectra in slit position S2 are extracted from 
the circumnuclear regions located to the North and North-West of the nucleus, 
and therefore any contribution from the underlying galaxy bulge is difficult
to assess. Slit position S1 crosses the galactic nucleus. This can be used to
estimate the underlying bulge contribution. For the blue spectra, it turns out to be 
almost negligible amounting to, at most, 10 per cent at the H$\beta$
line. For the red
spectra, the bulge contribution is more important. From Gaussian fits to the
$\lambda$\,8620\,\AA\ continuum profile of S1 we find it to be about 25 per
cent for R4, the 
weakest region, and the one closest to the nucleus.
On the other hand, the analysis of the broad near-IR HST-NICMOS images
shown in Fig.\ \ref{hst-slits} shows less contrast between the emission from
the 
regions and the underlying bulge which is very close to the image background
emission. Its contribution is about 25 per cent for
the weak regions in the central zone of NGC\,3310, that is in very good
agreement with the cluster identification made by
\cite{2002AJ....123.1381E} using the equivalent J and K-band HST-NICMOS
images and the ground base data from KPNO (J and K-bands).

Fig.\ \ref{spectra3310} shows the spectra of the observed circumnuclear
regions split into two panels corresponding to the blue and 
the red spectral ranges. The spectrum of the nucleus of NGC\,3310 is shown
in Fig.\ \ref{spectraN}.
The blue spectra show the Balmer H$\beta$
recombination line and the collisionally excited [O{\sc iii}] lines at
$\lambda\lambda$\,4959,\,5007\,\AA. Generally, these forbidden lines are very
weak (see for example Figs.\ 5 of Papers I and II), and, in some cases,
only the strongest $\lambda$ 5007 \AA\ is detected (right-hand panels of these
Figs.). However, in the case of
NGC\,3310 they are very strong, due to the low abundance of the
CNSFRs in this galaxy, with values between 0.2-0.4\,Z$_\odot$
\citep{1993MNRAS.260..177P}. These low
values of the abundances can be explained by the probably
unusual interaction history of the galaxy
\citep{2002AJ....123.1381E,2001A&A...376...59K,1996A&A...309..403M,1996ApJ...473L..21S,1981A&A....96..271B},
fuelling the ring with accreted neutral gas, as modeled by
\cite{1992MNRAS.259..345A} and \cite{1995ApJ...449..508P}. The red
spectra show the 
stellar CaT lines in absorption. In some cases these lines are contaminated by Paschen
emission which occurs at 
wavelengths very close to those of the CaT lines. Other emission features,
such as O{\sc i}\,$\lambda$\,8446, [Cl{\sc ii}]\,$\lambda$\,8579, Pa\,14 and
[Fe{\sc ii}]\,$\lambda$\,8617 are also present. 
In Fig.\ \ref{spectra3310} we can easily appreciate, for example in R4+R5, the Paschen
series from Pa13 to Pa22,  as well as the previously mentioned lines of O, Cl
and Fe. In all cases, a single Gaussian fit to the emission lines was
performed and the lines were subsequently subtracted 
\citep[see also][]{2004A&A...419L..43O,2008A&A...479..725C} after checking that the 
theoretically expected ratio between the Paschen lines was satisfied. The
observed red spectra are plotted with a dashed line.   
The solid line shows the subtracted spectra.


\begin{figure*}
\hspace{0.0cm}
\vspace{0.2cm}
\includegraphics[width=.48\textwidth,height=.30\textwidth]{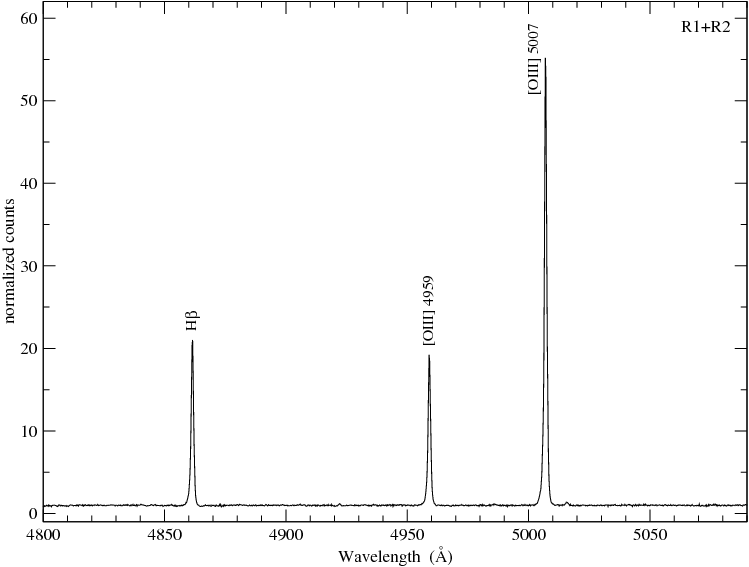}
\includegraphics[width=.48\textwidth,height=.30\textwidth]{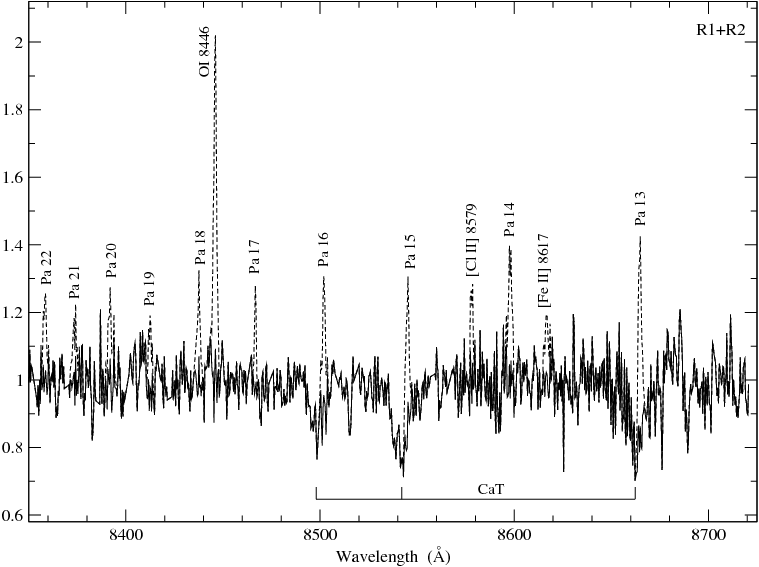}\\
\vspace{0.2cm}
\includegraphics[width=.48\textwidth,height=.30\textwidth]{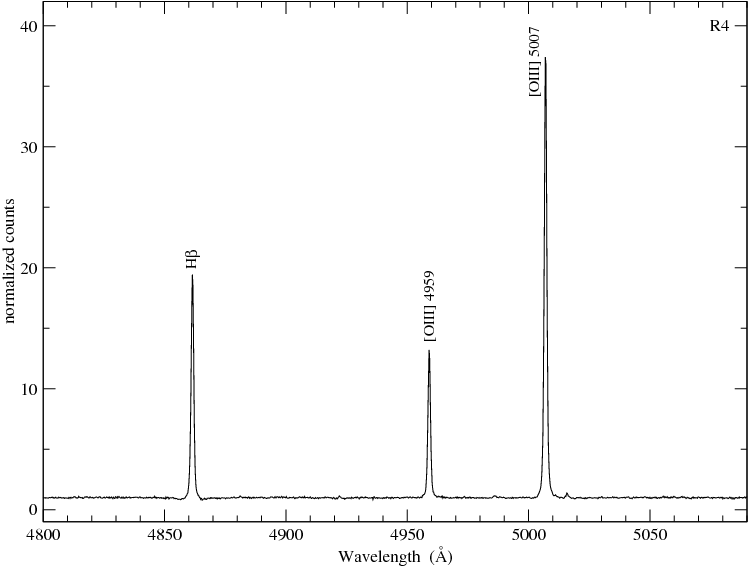}
\includegraphics[width=.48\textwidth,height=.30\textwidth]{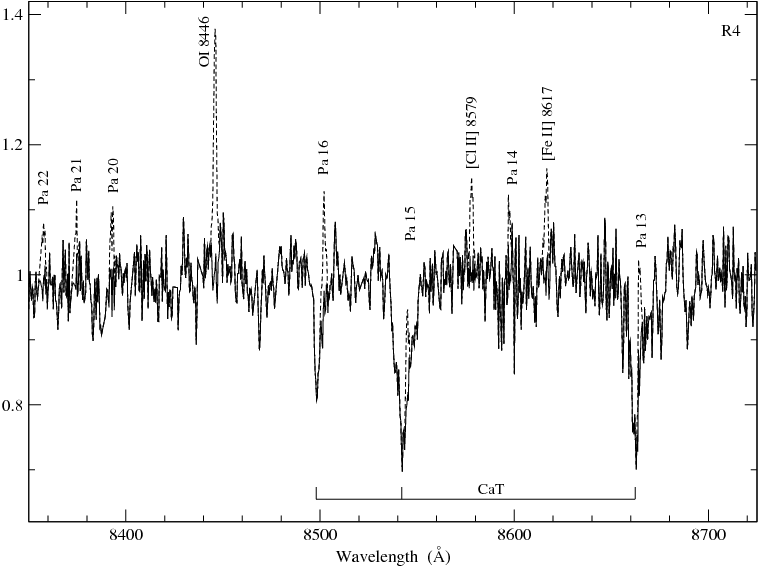}\\
\vspace{0.2cm}
\includegraphics[width=.48\textwidth,height=.30\textwidth]{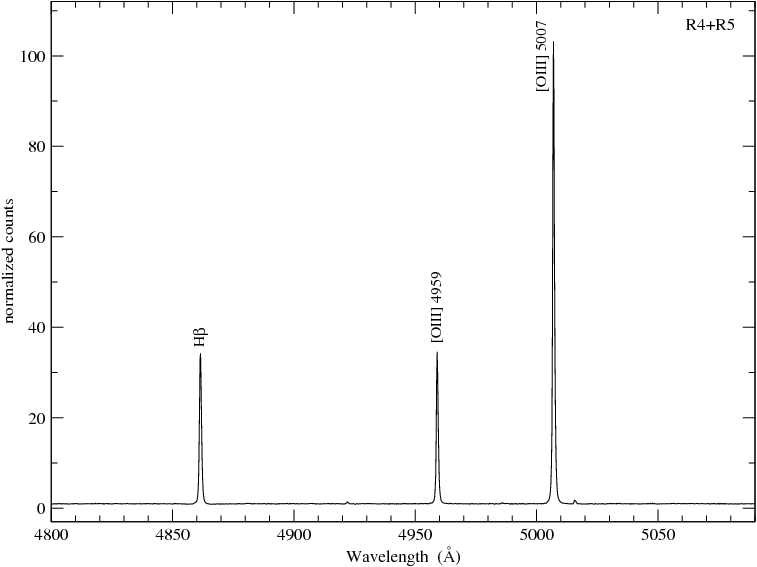}
\includegraphics[width=.48\textwidth,height=.30\textwidth]{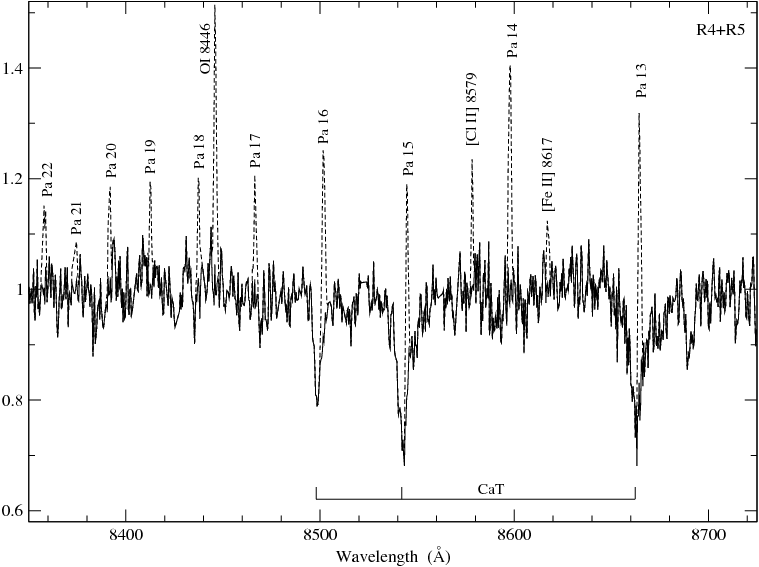}\\
\vspace{0.2cm}
\includegraphics[width=.48\textwidth,height=.30\textwidth]{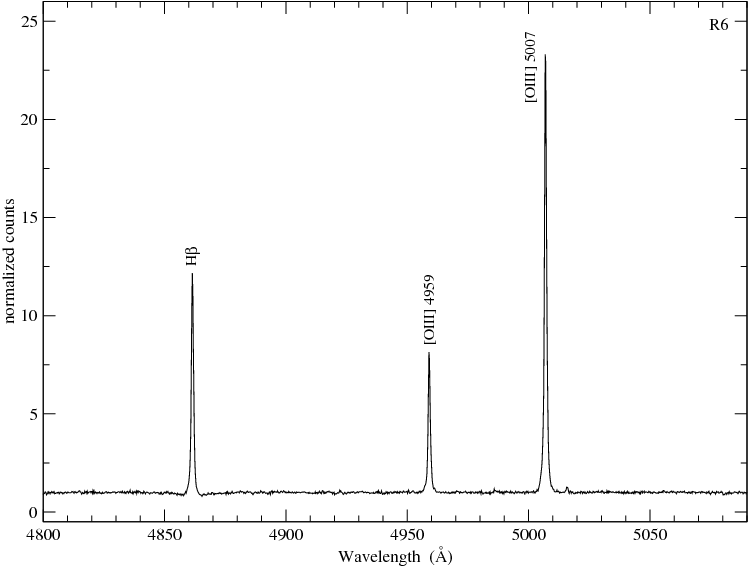}
\includegraphics[width=.48\textwidth,height=.30\textwidth]{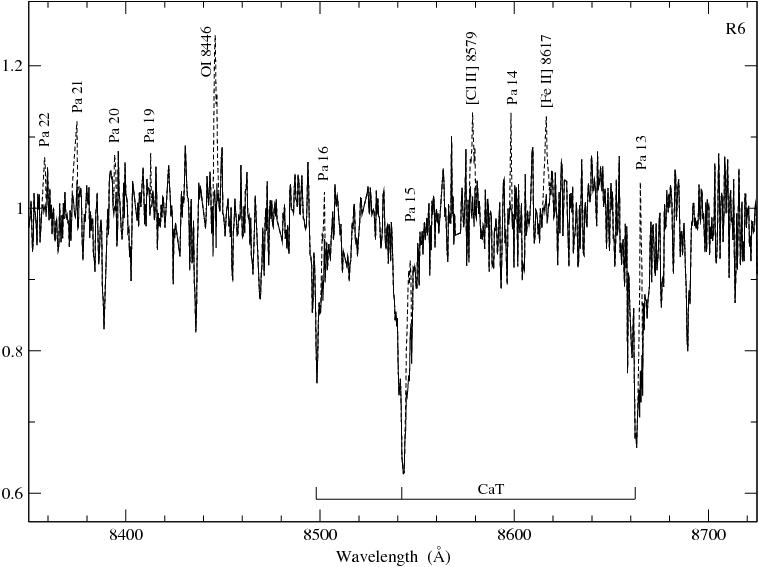}
\caption[]{Blue (left) and red (right) rest frame normalized spectra of the
  observed CNSFRs. For the red range, the dashed line shows the
  observed spectrum; the solid line represents the spectrum after subtracting
  the emission lines (see text).}
\label{spectra3310}
\end{figure*}

\setcounter{figure}{4}

\begin{figure*}
\hspace{0.0cm}
\vspace{0.2cm}
\includegraphics[width=.48\textwidth,height=.30\textwidth]{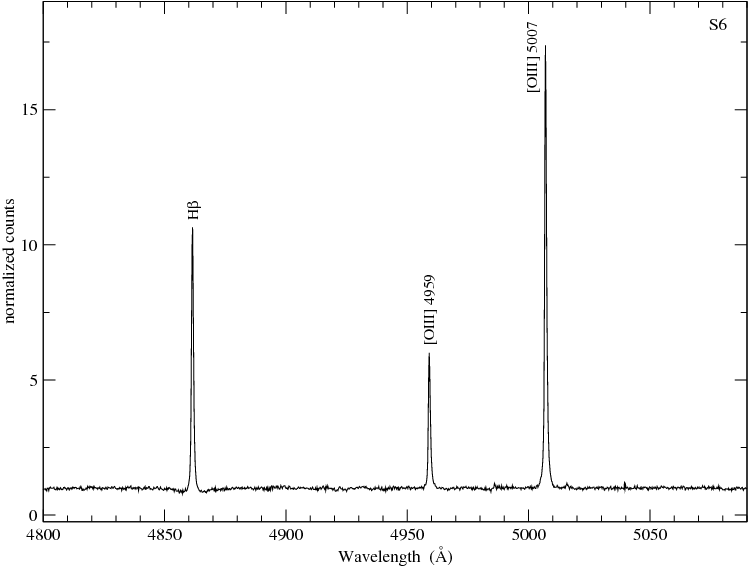}
\includegraphics[width=.48\textwidth,height=.30\textwidth]{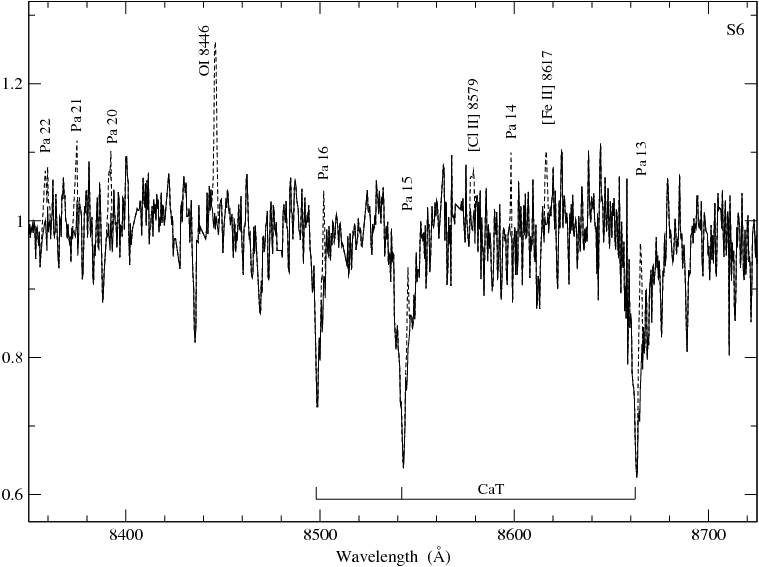}\\
\vspace{0.2cm}
\includegraphics[width=.48\textwidth,height=.30\textwidth]{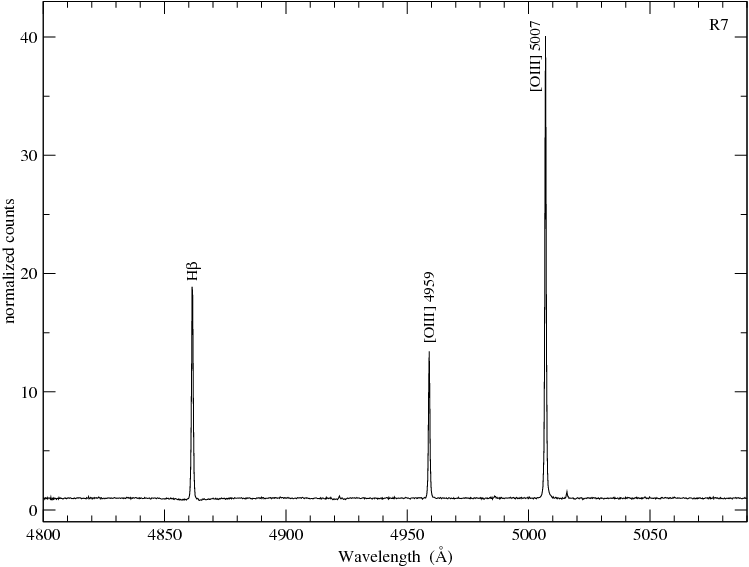}
\includegraphics[width=.48\textwidth,height=.30\textwidth]{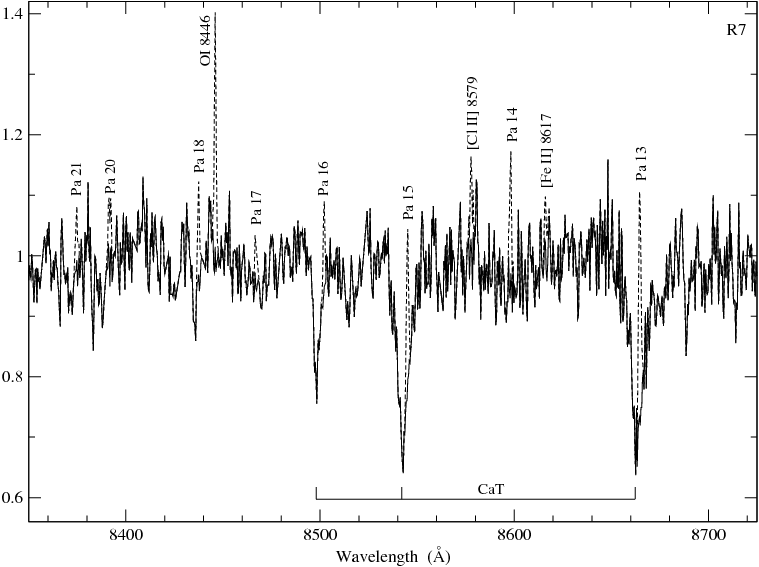}\\
\vspace{0.2cm}
\includegraphics[width=.48\textwidth,height=.30\textwidth]{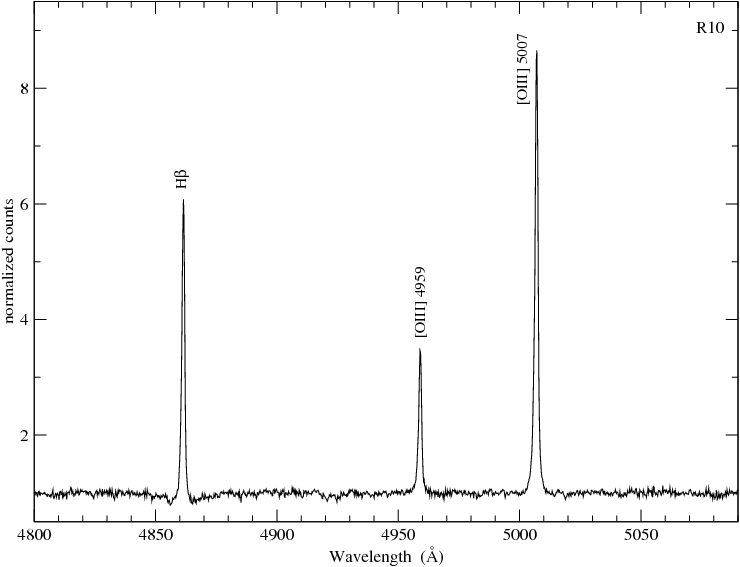}
\includegraphics[width=.48\textwidth,height=.30\textwidth]{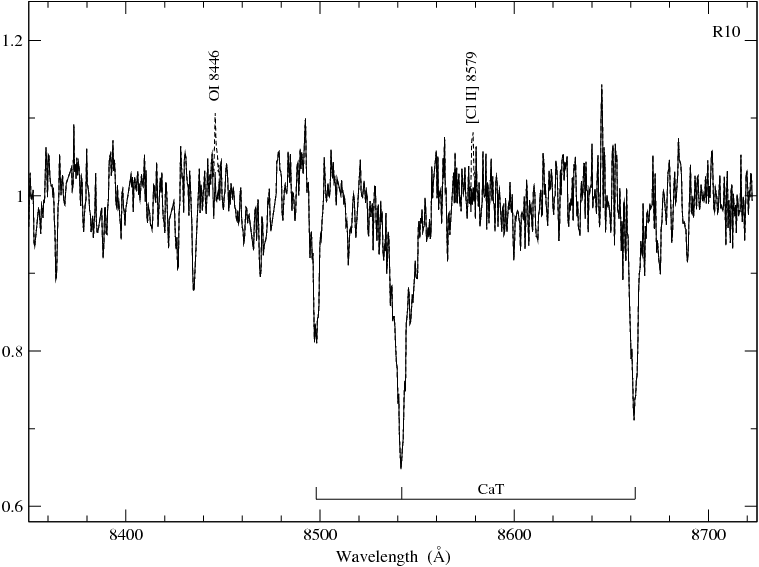}\\
\caption[]{({\it cont}) Blue (left) and red (right) rest frame normalized spectra of the
  observed CNSFRs. For the red range, the dashed line shows the
  observed spectrum; the solid line represents the spectrum after subtracting
  the emission lines (see text).} 
\label{spectra3310}
\end{figure*}


\begin{figure*}
\hspace{0.0cm}
\vspace{0.2cm}
\includegraphics[width=.48\textwidth,height=.30\textwidth]{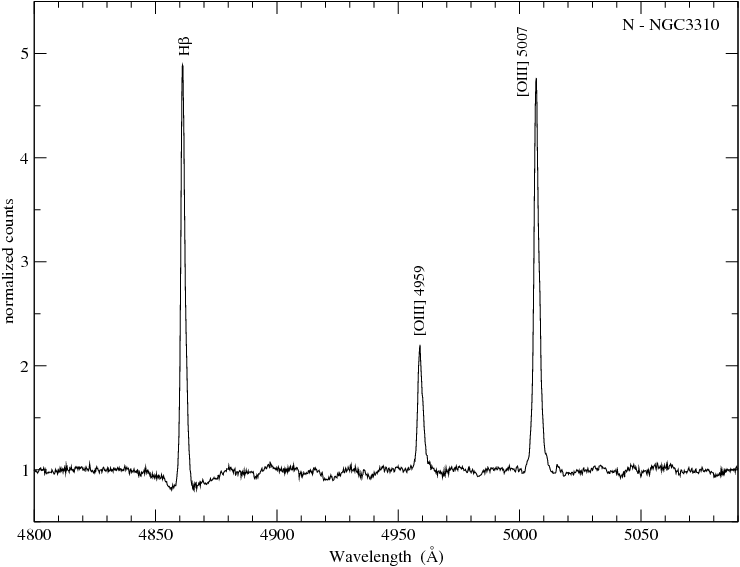}
\includegraphics[width=.48\textwidth,height=.30\textwidth]{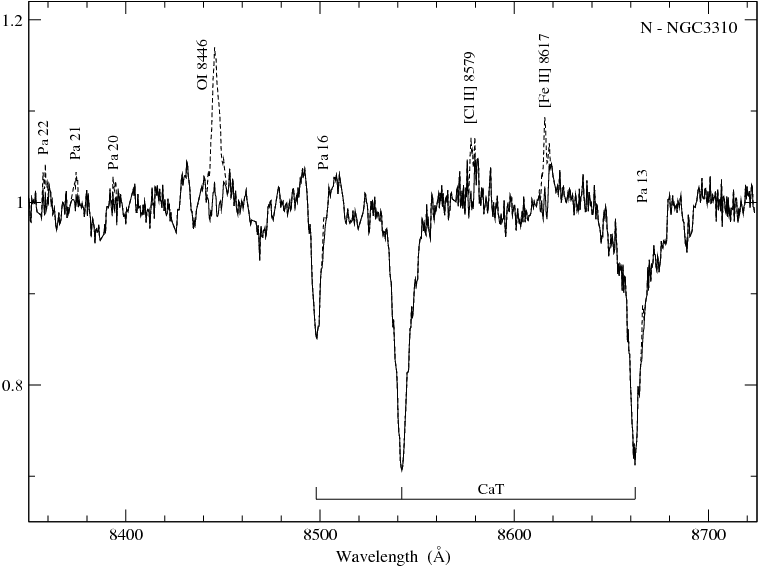}\\
\caption[]{Blue (left) and red (right) rest frame normalized spectra of the
  nucleus. In the red range, the dashed line shows the observed spectrum; the solid
  line represents the spectrum after subtracting the emission lines (see
  text).}
\label{spectraN}
\end{figure*}

\begin{figure*}
\hspace{0.0cm}
\vspace{0.2cm}
\includegraphics[width=.48\textwidth,height=.30\textwidth]{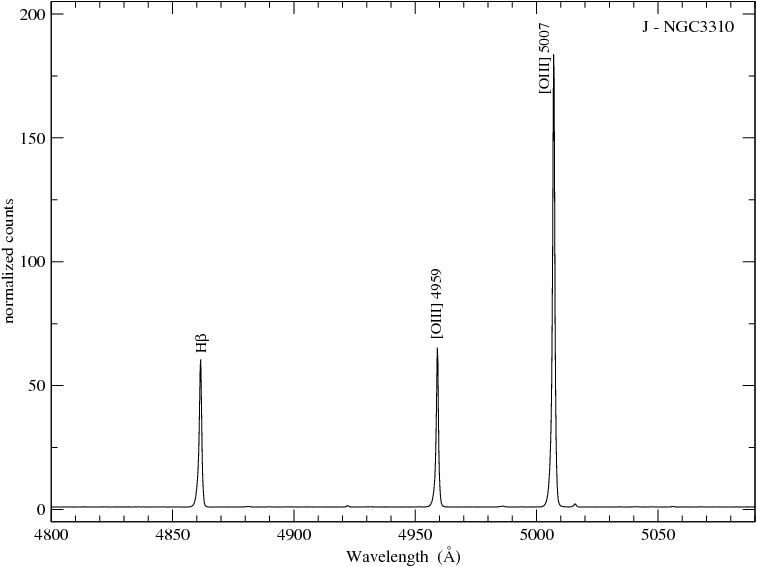}
\includegraphics[width=.48\textwidth,height=.30\textwidth]{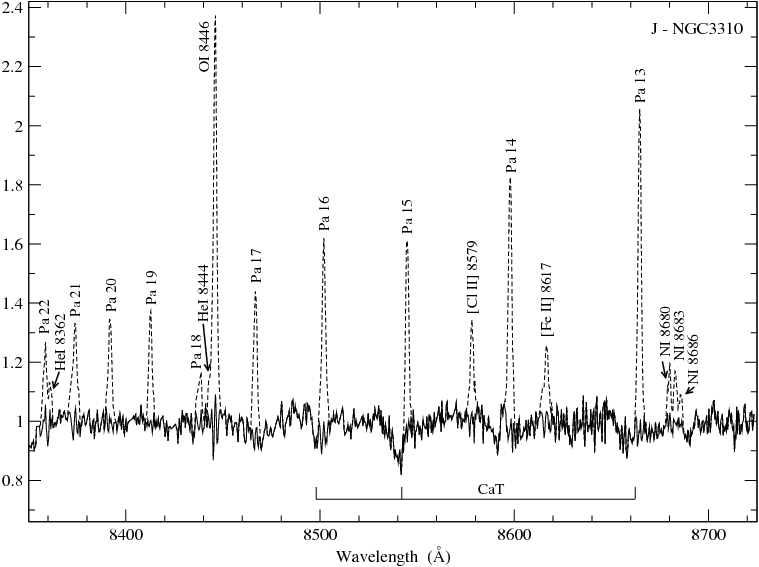}\\
\vspace{0.2cm}
\hspace{-0.485\textwidth}
\includegraphics[width=.48\textwidth,height=.30\textwidth]{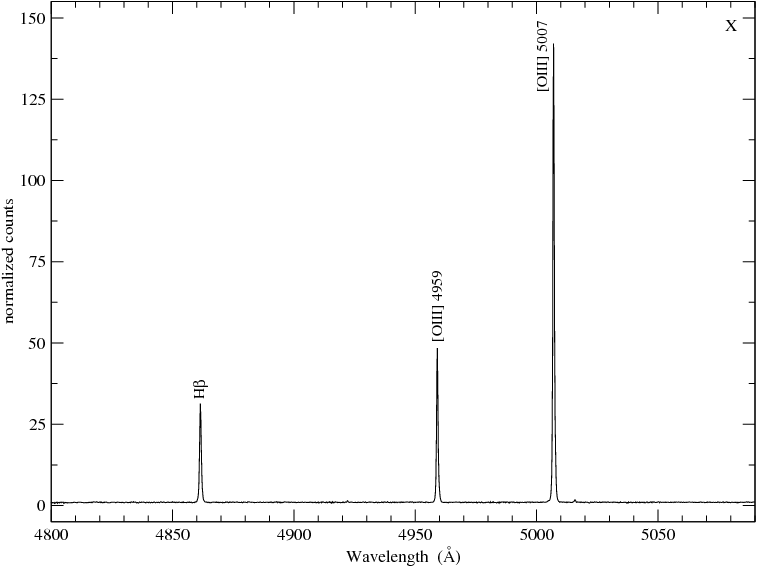}\\
\vspace{0.2cm}
\hspace{0.2cm}
\includegraphics[width=.48\textwidth,height=.30\textwidth]{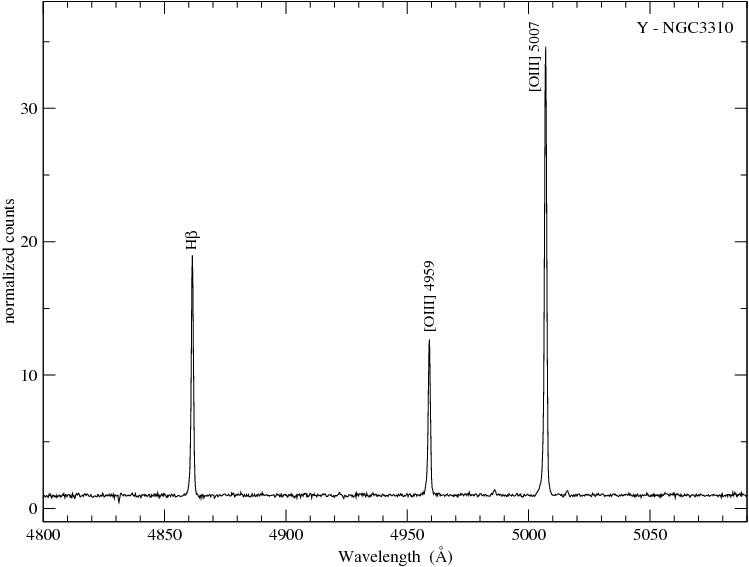}
\includegraphics[width=.48\textwidth,height=.30\textwidth]{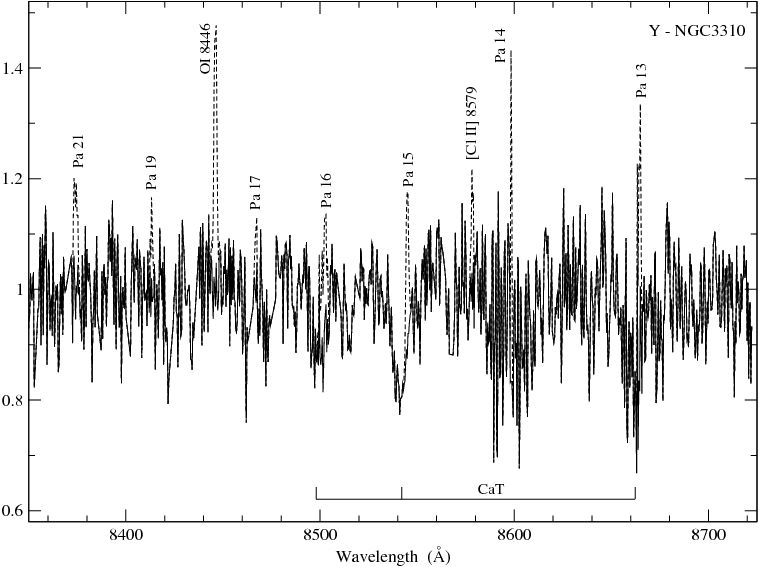}
\caption[]{Idem as Fig.\ \ref{spectra3310} for regions
  J, X (only blue) and Y. The dashed line shows the observed
  spectrum; the solid line represents the spectrum after subtracting the
  emission lines (see text).} 
\label{spectraemiknot}
\end{figure*}

Fig.\ \ref{spectraemiknot} shows the spectra of the almost pure emission
knots labelled X and Y, and those of the Jumbo
region. In all cases the blue range of the spectrum presents very
intense emission lines. 
The red spectral range presents a very weak and noisy continuum. In
the case of region X only noise is detected therefore no spectrum is shown in 
Fig.\ \ref{spectraemiknot}. In the other two regions we can
see a set of emission lines. The 
Jumbo region presents many strong lines, even He and N{\sc i} in
emission. Due to the low signal-to-noise ratio of the continuum and the
presence of the strong emission lines in these regions of NGC\,3310, we could
not obtain stellar spectra with enough signal in the CaT absorption
feature to allow an accurate measurement of velocity dispersions.


\subsection{Kinematics of stars and ionized gas} 
\label{Method}

A detailed description of the methods and techniques used to derive the values
of radial velocities and velocity dispersions as well as sizes, masses and
emission line fluxes has been given in Paper I. Therefore only a brief summary
is given below.  

\subsubsection*{Stellar analysis}

Stellar radial velocities and velocity dispersions were obtained from the
CaT absorption lines using the cross-correlation technique,
described in detail by \cite{1979AJ.....84.1511T}. 
This method requires the
comparison with a stellar template that represents the stellar population that
best reproduces the absorption features. This has been built from a set of 11
late-type giant and supergiant stars with strong CaT absorption lines. 
We have followed the work by \cite{1995ApJS...99...67N}
with the variation introduced by \cite{1997A&A...323..749P} of using the
individual stellar templates instead of an average. This procedure will allow
us to correct for the known possible mismatches between template stars and the
region composite spectrum. The implementation of the method in the 
external package of {\sc iraf xcsao} \citep{1998PASP..110..934K} has
been used. 

To determine the line-of-sight stellar velocity and velocity dispersion 
along each slit,  extractions 
were made every two pixels for slit position S1 and every three pixels for
slit position S2, with one pixel overlap between consecutive 
extractions in this latter case. In this way the S/N ratio and the spatial
resolution were optimized. Besides, the stellar velocity dispersion was
estimated at the position of each CNSFR and the nucleus using an aperture 
of five pixels in all cases, which corresponds to
1.0\,$\times$\,1.8\,arcsec$^2$. The velocity dispersion ($\sigma$) of the
stars ($\sigma_{\ast}$) is taken as the average of the $\sigma$ values found
for each 
stellar template, and its error is taken as the dispersion of the individual
values of $\sigma$ and the rms of the residuals of the wavelength fit. These
values  are listed in column 3 of Table \ref{disp} along with their
corresponding errors. 
Stellar velocity dispersions  of X, Y and the Jumbo region  could not be
estimated  due
to the low signal-to-noise ratio of the continua and the CaT absorption
features. The same is true for  region R1+R2, where the
red continuum and the CaT features after subtracting the emission lines have a
low signal-to-noise ratio, although  an estimate of the
stellar velocity dispersion could be given in this case.


\begin{table*}
\centering
{\footnotesize
\caption{Velocity dispersions.}
\begin{center}
\begin{tabular} {@{}l c c c c c c c c r@{}}
\hline
\hline
        &      &                  &  \multicolumn{2}{c}{{\it 1 component}} &
        \multicolumn{4}{c}{{\it 2 components}}    \\
        &      &           &       &      &  \multicolumn{2}{c}{{\it narrow}}
        & \multicolumn{2}{c}{{\it broad}} &   \\
 Region & Slit & $\sigma_{\ast}$  & $\sigma_{gas}$(H$\beta$) &
 $\sigma_{gas}$([O{\sc iii}]) &  $\sigma_{gas}$(H$\beta$)   &
        $\sigma_{gas}$([O{\sc iii}]) & $\sigma_{gas}$(H$\beta$)  &
        $\sigma_{gas}$([O{\sc iii}]) &  $\Delta$v$_{nb}$  \\
\hline

R1+R2 & S2 &   80:    & 33$\pm$4 & 31$\pm$4 & 24$\pm$5 & 22$\pm$5 & 54$\pm$7 & 50$\pm$4  & -10  \\
R4    & S1 & 36$\pm$3 & 34$\pm$3 & 32$\pm$3 & 28$\pm$4 & 26$\pm$3 & 55$\pm$5 & 52$\pm$4  & 10 \\
R4+R5 & S2 & 38$\pm$3 & 27$\pm$4 & 22$\pm$3 & 22$\pm$4 & 18$\pm$3 & 46$\pm$5 & 40$\pm$2  & 15 \\
R6    & S2 & 35$\pm$5 & 30$\pm$3 & 28$\pm$3 & 23$\pm$3 & 21$\pm$3 & 54$\pm$7 & 56$\pm$3  & 0   \\
S6    & S2 & 31$\pm$4 & 27$\pm$3 & 26$\pm$3 & 20$\pm$3 & 19$\pm$3 & 47$\pm$4 & 47$\pm$3  & 10 \\
R7    & S2 & 44$\pm$5 & 21$\pm$5 & 17$\pm$4 & 18$\pm$4 & 14$\pm$3 & 41$\pm$4 & 36$\pm$3  & 0   \\
R10   & S1 & 39$\pm$3 & 38$\pm$3 & 40$\pm$3 & 26$\pm$2 & 26$\pm$3 & 54$\pm$2 & 59$\pm$4  & -20  \\[2pt]
N     & S1 & 73$\pm$3 & 55$\pm$3 & 66$\pm$3 & 35$\pm$3 & 35$\pm$3 & 73$\pm$3 & 83$\pm$4  & 20 \\[2pt]
J     & S1 &    ---   & 34$\pm$2 & 30$\pm$2 & 25$\pm$2 & 22$\pm$2 & 61$\pm$3 & 57$\pm$3  & -25  \\
X     & S2 &    ---   & 22$\pm$4 & 18$\pm$3 & 18$\pm$4 & 14$\pm$3 & 40$\pm$5 & 30$\pm$3  & -5   \\
Y     & S1 &    ---   & 28$\pm$3 & 28$\pm$3 & 22$\pm$4 & 25$\pm$3 & 44$\pm$4 & 61$\pm$5  & 10 \\

\hline
\multicolumn{9}{@{}l}{velocity dispersions in km\,s$^{-1}$} \\

\end{tabular}
\end{center}
\label{disp}
}
\end{table*}


The radial velocities have been determined directly
from the position of the main peak of the cross-correlation of each galaxy
spectrum with each template in the rest frame. The average of these values is
the final adopted radial velocity.

\subsubsection*{Ionized gas analysis}

\begin{figure*}
\includegraphics[width=.48\textwidth,height=.30\textwidth,angle=0]{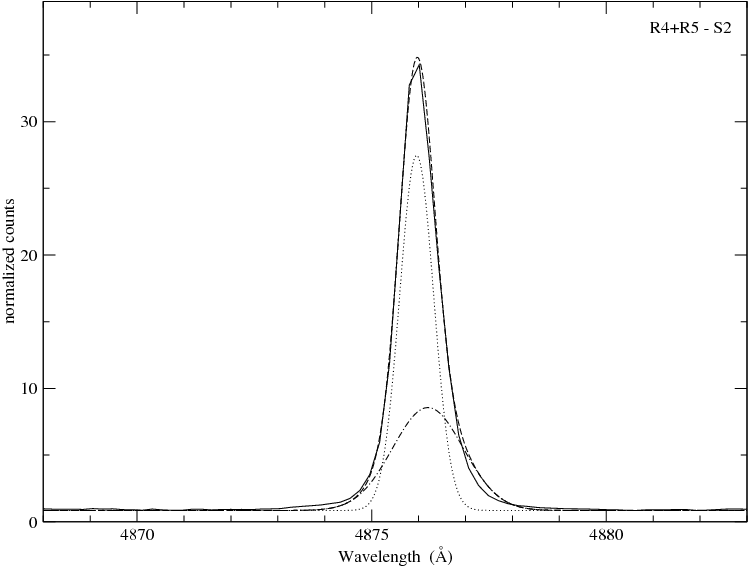}\hspace*{0.2cm}
\includegraphics[width=.48\textwidth,height=.30\textwidth,angle=0]{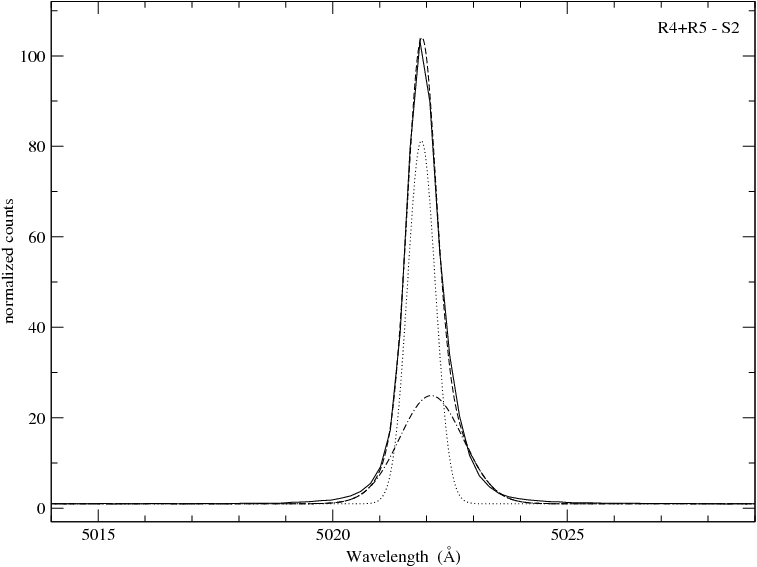}
\caption{Sections of the normalized spectrum of R4+R5 (solid line). The left
 panel shows from 4868 to 4883\,\AA\ 
 containing  H$\beta$ and the right panel shows from 5014 to 5029\,\AA\
 containing the [O{\sc iii}]\,$\lambda$\,5007\,\AA\ emission line. For both
 we have superposed the fits from the {\sc ngaussfit} task in {\sc iraf}; the
 dashed-dotted line is the broad component, the dotted line is the narrow
 component and the dashed line is the sum of both.}
\label{ngauss}
\end{figure*}

The velocity dispersion of the ionized gas was estimated for each observed 
CNSFR and for the galaxy nucleus from  Gaussian fits to the H$\beta$ 
and [O{\sc iii}]\,$\lambda$\,5007\,\AA\ emission lines using five pixel 
apertures, corresponding to 1.0\,$\times$\,1.9\,arcsec$^2$. For a single
Gaussian fit,  the position and width of a given emission line is taken as the
average of the fitted Gaussians to the whole line using three different
suitable continua \citep{2000MNRAS.317..907J}, and their errors are given by  
the dispersion of these measurements taking into account the rms of the
wavelength calibration.
In all the studied regions, however, the best fit for the
emission lines are obtained with two different components having different 
radial velocities of up to 25 km s$ ^{-1} $. The radial velocities found for
the narrow and broad components of both H$\beta$ and [O{\sc iii}], are the
same within the errors. An example of the two-Gaussian fit for R4+R5 in S2 is
shown in Fig.\ \ref{ngauss}.

For each CNSFR, the gas velocity dispersion for the H$\beta$ and [O{\sc
iii}]\,$\lambda$\,5007\,\AA\ lines derived using single and double line
Gaussian fits, and their corresponding errors are listed in Table
\ref{disp}. Columns 4 and 5, labelled `One component', give the results for
the single Gaussian fit. Columns 6 and 7, and 8 and 9,  labelled `Two
components - Narrow' and `Two components - Broad', respectively,  list the
results for the two component fits. The last column of the table, labelled
$\Delta$v$_{nb}$, gives the velocity difference between the narrow and broad
components. This is calculated as the average of the H$\beta$ and [O{\sc iii}]
fit differences. Taking into account the errors in the two component fits, the
errors in these velocity differences vary from  5 to 10\,km\,s$^{-1}$.

We have also determined the distribution along each slit position of the
radial velocities and the velocity dispersions of the ionized gas using the
same procedure as for the stars, that is using spectra extracted
every two pixels for S1 and every three pixels, superposing one pixel for
consecutive extractions, for S2. These spectra, however, do not have the
required S/N ratio to allow an acceptable two-component fit, therefore a
single-Gaussian component has been used. The goodness of this procedure is
discussed in Section 4 below.

\subsection{Emission line ratios}
\label{lineratios}

\begin{figure*}
\includegraphics[width=.48\textwidth,height=.30\textwidth,angle=0]{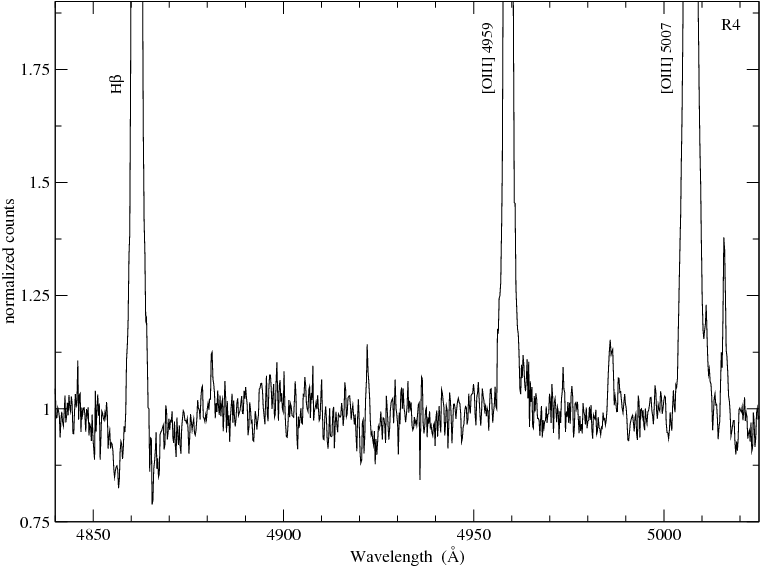}
\includegraphics[width=.48\textwidth,height=.30\textwidth,angle=0]{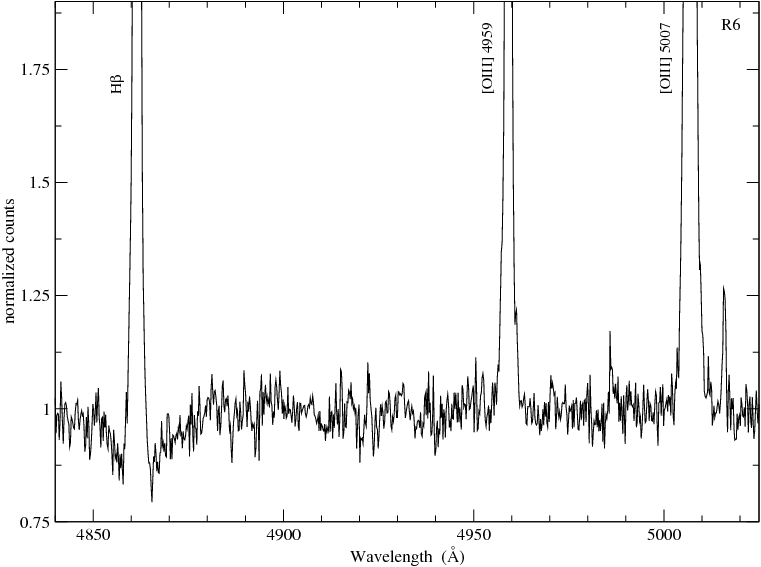}
\caption{Enlargement of the blue rest frame normalized spectra of R4 (left) and R6 (right).}
\label{enlarg}
\end{figure*}

We have used two different ways to integrate the intensity of a given line:
(1) if an adequate fit was attained by a single Gaussian, the emission line 
intensities were measured using the {\sc splot} task in {\sc iraf}. For the H$\beta$
emission lines a conspicuous underlying stellar population is inferred
from the presence of absorption features that depress the lines
\cite[e.g.~see discussion in][]{1988MNRAS.231...57D}. Examples of this effect
can be appreciated in Fig.\ \ref{enlarg}. We have defined a pseudo-continuum 
at the base  of the line to measure the line intensities and minimize
the errors introduced by the underlying population \cite[for details
  see][]{2006MNRAS.372..293H}. (2) When the optimal fit was obtained by two
Gaussians the individual intensities  of the narrow and broad components
are estimated from the fitting parameters
(I\,=\,1.0645\,A\,$\times$\,FWHM\,=\,$\sqrt{2\pi}$\,A\,$\sigma$; where I is
the Gaussian intensity, A is the amplitude of the Gaussian, FWHM is the full
width at half-maximum and $\sigma$ is the dispersion of the Gaussian). 
A pseudo-continuum for the H$\beta$ emission line was also defined in these
cases. The statistical errors associated with the observed
emission fluxes have been calculated with the expression
$\sigma_{l}$\,=\,$\sigma_{c}$N$^{1/2}$[1 + EW/(N$\Delta$)]$^{1/2}$, where
$\sigma_{l}$ is  the error in the observed line flux, $\sigma_{c}$ represents
the standard 
deviation in a box near the measured emission line and stands for the error in
the continuum placement, N is the number of pixels used in the measurement of 
the line intensity, EW is the line equivalent width and $\Delta$ is 
the wavelength dispersion in \AA\,pixel$^{-1}$ \citep{1994ApJ...437..239G}. 
For the H$\beta$ emission line we have
doubled the derived error, $\sigma_{l}$, in order to take into account the
uncertainties introduced by the presence of the underlying stellar population 
\citep{2006MNRAS.372..293H}. 

The logarithmic ratio between the emission line intensities of [O{\sc
iii}]\,$\lambda$\,5007\,\AA\ and H$\beta$ and their corresponding errors
are presented in Table \ref{ratios}. We have also listed the
logarithmic ratio between the emission line fluxes of [N{\sc
ii}]\,$\lambda$\,6584\,\AA\ and H$\alpha$ together with their corresponding
errors. These values listed for R4, R4+R5, S6, J and the
nucleus are from Pastoriza et al.\ (1993; S6 seems to be their region L). For
the rest of the regions we  
used an extrapolation of the results given by them from the spatial profiles of
these emission lines. They reported a constant value of 0.2 for the [N{\sc
ii}]\,/\,H$\alpha$ ratio for the \HII\ regions in their slit positions 2 and
3, while this value changes from 0.2 to 0.5 along position 1 
over the nucleus of the galaxy, being 0.23 and 0.27 for regions B and L,
respectively. We adopt for R1+R2, R6, R7, R10, X and Y, a value of 0.2
without assigning any error to it.



\begin{table*}
\centering
\caption[]{Line ratios.}
\begin{tabular} {l c c c c c}
\hline
        &      &         One component &
        \multicolumn{2}{c}{Two components}   & \\
        &      &    &  Narrow  & Broad &  \\
 Region & Slit & log([O{\sc iii}]5007/H$\beta$) & log([O{\sc
        iii}]5007/H$\beta$) & log([O{\sc iii}]5007/H$\beta$) & log([N{\sc
        ii}]6584/H$\alpha$)$^a$ \\
\hline

R1+R2 &  S2   &   0.42$\pm$0.01   &  0.39$\pm$0.01  &  0.45$\pm$0.03 &  -0.70: \\
R4    &  S1   &   0.28$\pm$0.01   &  0.28$\pm$0.01  &  0.28$\pm$0.04 &  -0.69$\pm$0.01 \\
R4+R5 &  S2   &   0.42$\pm$0.01   &  0.39$\pm$0.01  &  0.44$\pm$0.03 &  -0.69$\pm$0.01 \\
R6    &  S2   &   0.28$\pm$0.01   &  0.30$\pm$0.01  &  0.28$\pm$0.06 &  -0.70: \\
S6    &  S2   &   0.21$\pm$0.01   &  0.21$\pm$0.01  &  0.22$\pm$0.08 &  -0.56$\pm$0.01 \\
R7    &  S2   &   0.23$\pm$0.01   &  0.21$\pm$0.01  &  0.32$\pm$0.09 &  -0.70: \\
R10   &  S1   &   0.19$\pm$0.01   &  0.20$\pm$0.04  &  0.20$\pm$0.08 &  -0.70: \\[2pt] 
N     &  S2   &   0.04$\pm$0.02   & -0.13$\pm$0.06  &  0.11$\pm$0.07 &  -0.30$\pm$0.01 \\[2pt]
J     &  S1   &   0.45$\pm$0.01   &  0.45$\pm$0.01  &  0.45$\pm$0.01 &  -0.80$\pm$0.01 \\
X     &  S2   &   0.60$\pm$0.01   &  0.55$\pm$0.01  &  0.67$\pm$0.04 &  -0.70: \\
Y     &  S1   &   0.28$\pm$0.01   &  0.43$\pm$0.01  &  0.00$\pm$0.04 &  -0.70: \\

\hline

\multicolumn{6}{@{}l}{$^a$From Pastoriza et al.\ (1993).}

\end{tabular}
\label{ratios}
\end{table*}


\section{Dynamical mass derivation}

The mass of a virialized stellar system is given by two parameters: its
velocity dispersion and its size.

In order to determine the sizes of the stellar clusters within our observed
CNSFRs, we have used the retrieved wide V HST image which provides a spatial
resolution of 0.045\,\arcsec\ per pixel. At the distance of NGC\,3310, this
resolution corresponds to 3.3\,pc/pixel. Fig. \ref{sizes} shows enlargements
around the different regions studied with intensity contours overlapped.  

We find, as expected, that the eight CNSFRs studied here are formed by a large
number of individual star-forming clusters. When we analyze the F606W-PC1
image of NGC\,3310, we find another region very close to R6, not classified 
by \cite{2000MNRAS.311..120D}, probably due to the lower spatial resolution
of the data used in that work, and labelled S6 by us, which seems to be
coincident with 
knot L of \cite{1993MNRAS.260..177P}. For all the regions of this
galaxy we find a principal knot and several secondary knots with lower peak
intensities, except for R1 and R2, which present only one and two secondary
knots, respectively. The same criterion as in Papers I and II has been
applied to name the regions. For the other regions of this galaxy, R4, R5, R6,
S6, R7 and R10, we find 31, 8, 14, 9, 29 and 15 individual clusters,
respectively. In all cases the knots have been found with a detection
level of 10\,$\sigma$ above the background. All these knots are within the
radius of the regions defined by \cite{2000MNRAS.311..120D}, except for S6.
We have to remark that our search for knots has not been exhaustive since that
is not the aim of this work.

\begin{figure*}
\centering
\begin{minipage}[c]{6.1in}
\includegraphics[width=.61\textwidth,angle=90]{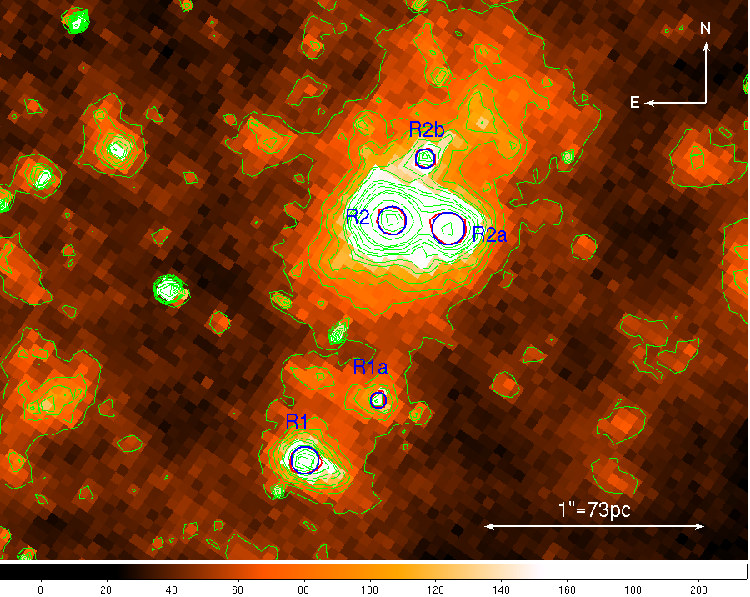}
\includegraphics[width=.61\textwidth,angle=90]{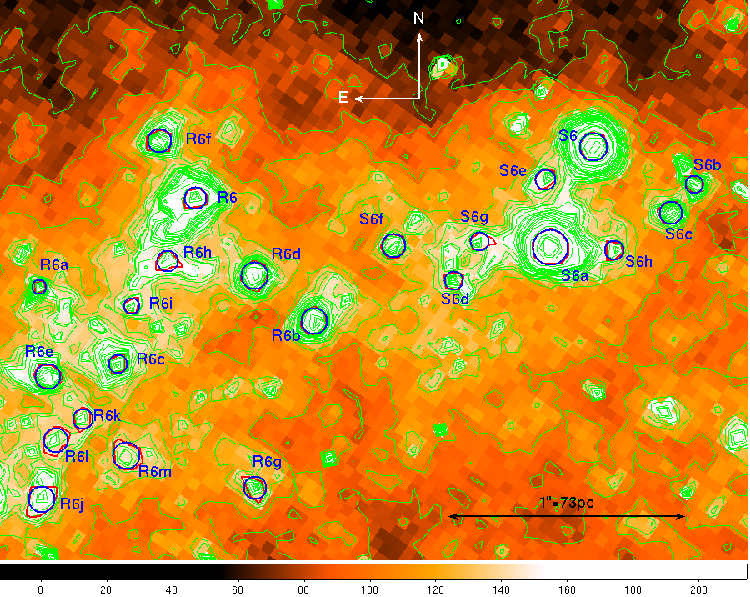}\\
\includegraphics[width=.61\textwidth,angle=90]{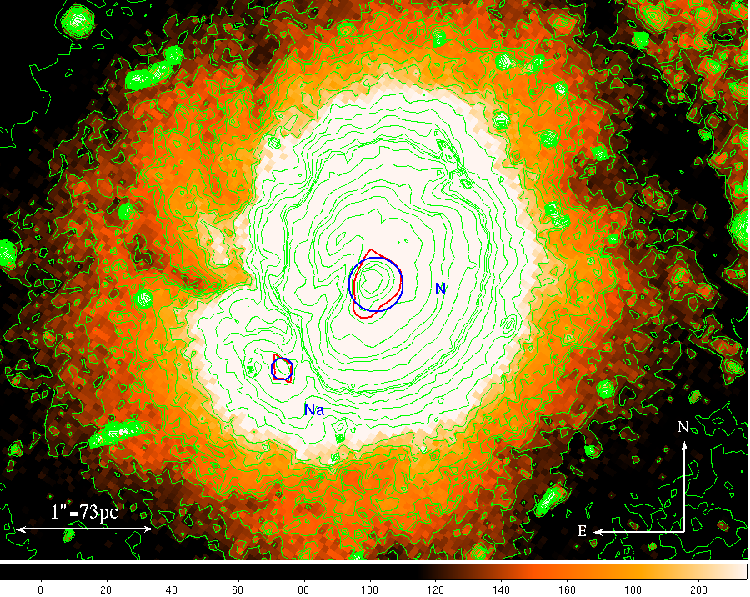}
\includegraphics[width=.61\textwidth,angle=90]{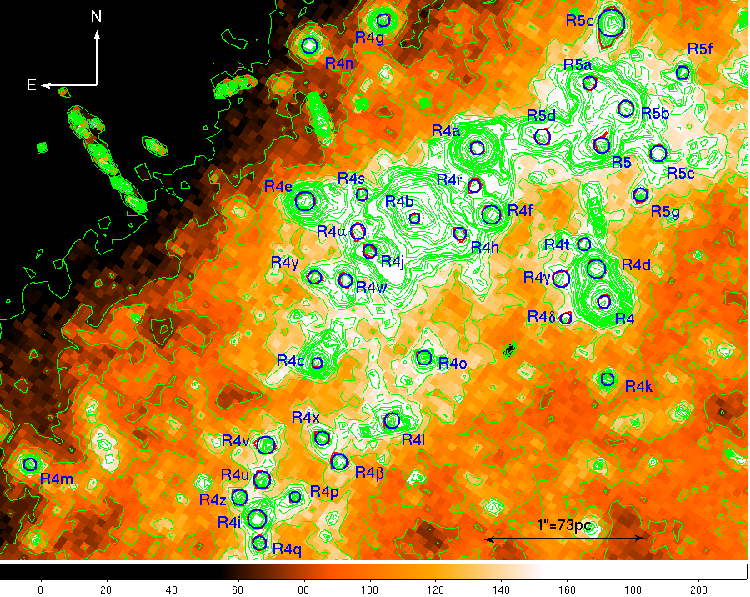}\\
\end{minipage}
\begin{minipage}[c]{0.5in}
\rotcaption{Enlargements of the F606W image around the nucleus and the CNSFRs
  of our study with the contours overlapped. The circles correspond to the
  adopted radius for each region. 
[{\it See the electronic edition of the Journal for a  
  colour version of this figure where the adopted radii are in blue and
  the contours corresponding to the half light brightness are in red.}]} 
\label{sizes}
\end{minipage}
\end{figure*}

\setcounter{figure}{9}

\begin{figure}
\centering
\includegraphics[width=0.60\textwidth,angle=90]{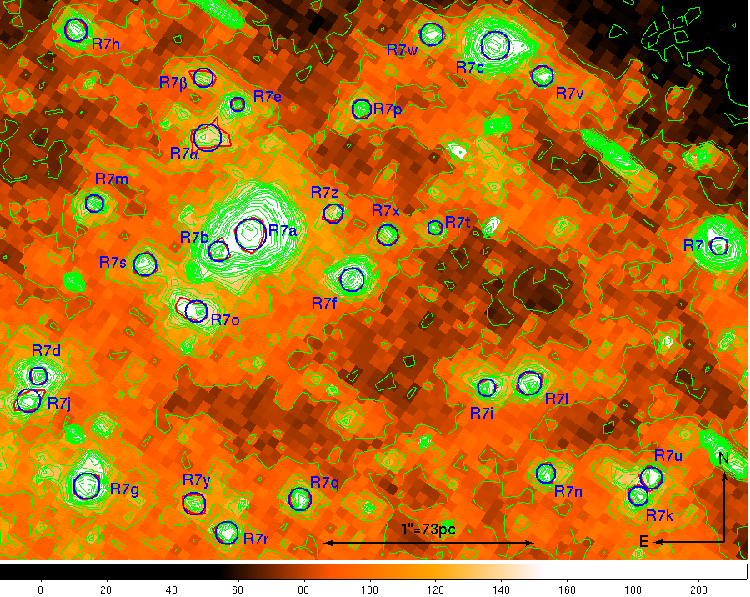}\\
\includegraphics[width=0.60\textwidth,angle=90]{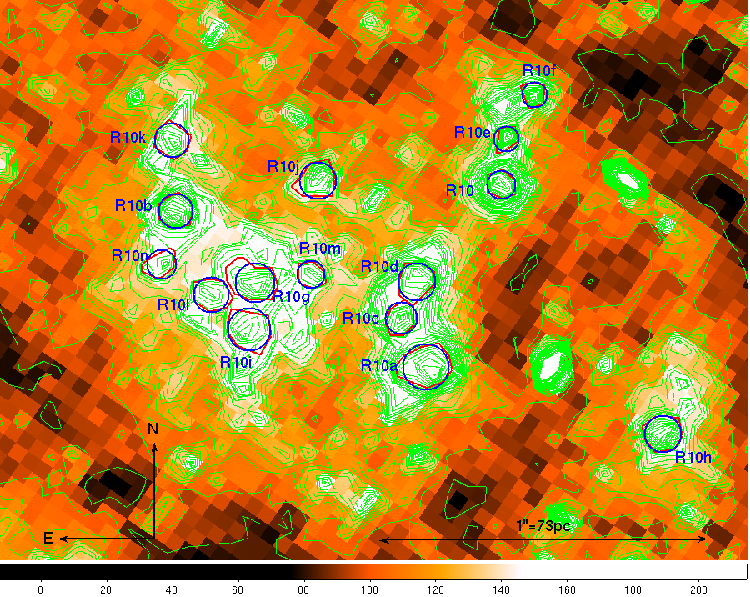}
\caption{({\it cont}) Enlargements of the F606W image around the CNSFRs of our
  study with the contours overlapped. The circles correspond to the adopted
  radius for each region. 
[{\it See the electronic edition of the Journal for a  
  colour version of this figure where the adopted radii are in blue and
  the contours corresponding to the half light brightness are in red.}]}
\label{sizes}
\end{figure}


We have fitted circular regions to the intensity contours corresponding to the
half light brightness distribution of each single structure (see
Fig. \ref{sizes}), following the procedure given in
\cite{1995AJ....110.2665M}, assuming that the regions have a circularly
symmetric Gaussian profile. The radii of the single knots vary between 2.2 and
6.2\,pc. Table \ref{knots} gives, for each identified knot, the position, as
given by the astrometric calibration of the HST image; the radius of the
circular region defined as described above together with its error;  and the
peak intensity in counts, as measured from the WFPC2 image. For
the nucleus of NGC\,3310 we measure a radius of 14.2\,pc. 


\begin{table*}
\centering
\caption[]{Positions, radii and peak intensities derived from the HST-F606W image.}
\begin{tabular}{@{}lcccc@{}p{0.6cm}@{}lcccc@{}}
\cline{1-5}\cline{7-11} 
Region &  \multicolumn{2}{c}{position} &  R & I & & Region &  \multicolumn{2}{c}{position} &  R & I \\
  &  $\alpha_{J2000.0}$ & $\delta_{J2000.0}$ & (pc) & (counts) & & &  $\alpha_{J2000.0}$ & $\delta_{J2000.0}$ & (pc) & (counts) \\
\cline{1-5}\cline{7-11} 
 
R1	 & 10$^{\romn h}$38$^{\romn m}$47\fs070 & +53$^\circ$30\arcmin14\farcs46 & 4.6$\pm$0.3 & 322  & &  S6	 & 10$^{\romn h}$38$^{\romn m}$45\fs748 & +53$^\circ$30\arcmin18\farcs42 & 4.2$\pm$0.2 & 544 \\
R1a	 & 10$^{\romn h}$38$^{\romn m}$47\fs032 & +53$^\circ$30\arcmin14\farcs74 & 2.6$\pm$0.2 & 179  & &  S6a	 & 10$^{\romn h}$38$^{\romn m}$45\fs768 & +53$^\circ$30\arcmin18\farcs00 & 5.5$\pm$0.2 & 385 \\
R2	 & 10$^{\romn h}$38$^{\romn m}$47\fs026 & +53$^\circ$30\arcmin15\farcs55 & 4.7$\pm$0.2 & 1271 & &  S6b	 & 10$^{\romn h}$38$^{\romn m}$45\fs700 & +53$^\circ$30\arcmin18\farcs26 & 2.6$\pm$0.3 & 237 \\
R2a	 & 10$^{\romn h}$38$^{\romn m}$46\fs997 & +53$^\circ$30\arcmin15\farcs52 & 5.5$\pm$0.5 & 515  & &  S6c	 & 10$^{\romn h}$38$^{\romn m}$45\fs711 & +53$^\circ$30\arcmin18\farcs14 & 3.4$\pm$0.3 & 204 \\
R2b	 & 10$^{\romn h}$38$^{\romn m}$47\fs008 & +53$^\circ$30\arcmin15\farcs84 & 3.3$\pm$0.2 & 287  & &  S6d	 & 10$^{\romn h}$38$^{\romn m}$45\fs813 & +53$^\circ$30\arcmin17\farcs86 & 2.7$\pm$0.3 & 180 \\
R4	 & 10$^{\romn h}$38$^{\romn m}$46\fs109 & +53$^\circ$30\arcmin16\farcs04 & 2.9$\pm$0.2 & 2409 & &  S6e	 & 10$^{\romn h}$38$^{\romn m}$45\fs770 & +53$^\circ$30\arcmin18\farcs28 & 3.0$\pm$0.2 & 174 \\
R4a	 & 10$^{\romn h}$38$^{\romn m}$46\fs198 & +53$^\circ$30\arcmin17\farcs01 & 3.1$\pm$0.2 & 1942 & &  S6f	 & 10$^{\romn h}$38$^{\romn m}$45\fs842 & +53$^\circ$30\arcmin18\farcs00 & 3.6$\pm$0.2 & 173 \\
R4b	 & 10$^{\romn h}$38$^{\romn m}$46\fs242 & +53$^\circ$30\arcmin16\farcs57 & 2.5$\pm$0.2 & 1038 & &  S6g	 & 10$^{\romn h}$38$^{\romn m}$45\fs801 & +53$^\circ$30\arcmin18\farcs02 & 2.9$\pm$0.3 & 172 \\
R4c	 & 10$^{\romn h}$38$^{\romn m}$46\fs311 & +53$^\circ$30\arcmin15\farcs66 & 2.4$\pm$0.2 & 584  & &  S6h	 & 10$^{\romn h}$38$^{\romn m}$45\fs738 & +53$^\circ$30\arcmin17\farcs98 & 2.8$\pm$0.3 & 166 \\
R4d	 & 10$^{\romn h}$38$^{\romn m}$46\fs115 & +53$^\circ$30\arcmin16\farcs25 & 4.1$\pm$0.2 & 494  & &  R7	 & 10$^{\romn h}$38$^{\romn m}$45\fs270 & +53$^\circ$30\arcmin18\farcs18 & 3.1$\pm$0.2 & 639 \\
R4e	 & 10$^{\romn h}$38$^{\romn m}$46\fs319 & +53$^\circ$30\arcmin16\farcs67 & 4.2$\pm$0.3 & 489  & &  R7a	 & 10$^{\romn h}$38$^{\romn m}$45\fs520 & +53$^\circ$30\arcmin18\farcs24 & 5.3$\pm$0.2 & 599 \\
R4f	 & 10$^{\romn h}$38$^{\romn m}$46\fs188 & +53$^\circ$30\arcmin16\farcs59 & 4.2$\pm$0.2 & 403  & &  R7b	 & 10$^{\romn h}$38$^{\romn m}$45\fs537 & +53$^\circ$30\arcmin18\farcs16 & 3.5$\pm$0.3 & 449 \\
R4g	 & 10$^{\romn h}$38$^{\romn m}$46\fs264 & +53$^\circ$30\arcmin17\farcs81 & 2.9$\pm$0.2 & 391  & &  R7c	 & 10$^{\romn h}$38$^{\romn m}$45\fs390 & +53$^\circ$30\arcmin19\farcs14 & 4.8$\pm$0.2 & 390 \\
R4h	 & 10$^{\romn h}$38$^{\romn m}$46\fs210 & +53$^\circ$30\arcmin16\farcs47 & 2.8$\pm$0.4 & 336  & &  R7d	 & 10$^{\romn h}$38$^{\romn m}$45\fs633 & +53$^\circ$30\arcmin17\farcs57 & 3.0$\pm$0.2 & 275 \\
R4i	 & 10$^{\romn h}$38$^{\romn m}$46\fs353 & +53$^\circ$30\arcmin14\farcs68 & 4.2$\pm$0.2 & 304  & &  R7e	 & 10$^{\romn h}$38$^{\romn m}$45\fs527 & +53$^\circ$30\arcmin18\farcs86 & 2.4$\pm$0.3 & 273 \\
R4j	 & 10$^{\romn h}$38$^{\romn m}$46\fs274 & +53$^\circ$30\arcmin16\farcs36 & 3.0$\pm$0.3 & 284  & &  R7f	 & 10$^{\romn h}$38$^{\romn m}$45\fs466 & +53$^\circ$30\arcmin18\farcs03 & 3.9$\pm$0.2 & 264 \\
R4k	 & 10$^{\romn h}$38$^{\romn m}$46\fs107 & +53$^\circ$30\arcmin15\farcs56 & 2.7$\pm$0.2 & 271  & &  R7g	 & 10$^{\romn h}$38$^{\romn m}$45\fs607 & +53$^\circ$30\arcmin17\farcs05 & 4.5$\pm$0.2 & 231 \\
R4l	 & 10$^{\romn h}$38$^{\romn m}$46\fs258 & +53$^\circ$30\arcmin15\farcs30 & 3.4$\pm$0.5 & 268  & &  R7h	 & 10$^{\romn h}$38$^{\romn m}$45\fs613 & +53$^\circ$30\arcmin19\farcs21 & 3.9$\pm$0.2 & 209 \\
R4m	 & 10$^{\romn h}$38$^{\romn m}$46\fs513 & +53$^\circ$30\arcmin15\farcs02 & 2.8$\pm$0.3 & 259  & &  R7i	 & 10$^{\romn h}$38$^{\romn m}$45\fs394 & +53$^\circ$30\arcmin17\farcs51 & 3.0$\pm$0.2 & 201 \\
R4n	 & 10$^{\romn h}$38$^{\romn m}$46\fs316 & +53$^\circ$30\arcmin17\farcs66 & 3.4$\pm$0.3 & 250  & &  R7j	 & 10$^{\romn h}$38$^{\romn m}$45\fs638 & +53$^\circ$30\arcmin17\farcs45 & 4.0$\pm$0.5 & 200 \\
R4o	 & 10$^{\romn h}$38$^{\romn m}$46\fs236 & +53$^\circ$30\arcmin15\farcs69 & 3.1$\pm$0.2 & 241  & &  R7k	 & 10$^{\romn h}$38$^{\romn m}$45\fs313 & +53$^\circ$30\arcmin17\farcs00 & 3.4$\pm$0.3 & 198 \\
R4p	 & 10$^{\romn h}$38$^{\romn m}$46\fs327 & +53$^\circ$30\arcmin14\farcs82 & 2.4$\pm$0.2 & 241  & &  R7l	 & 10$^{\romn h}$38$^{\romn m}$45\fs371 & +53$^\circ$30\arcmin17\farcs53 & 4.1$\pm$0.3 & 191 \\
R4q	 & 10$^{\romn h}$38$^{\romn m}$46\fs352 & +53$^\circ$30\arcmin14\farcs53 & 3.1$\pm$0.4 & 227  & &  R7m	 & 10$^{\romn h}$38$^{\romn m}$45\fs603 & +53$^\circ$30\arcmin18\farcs39 & 3.0$\pm$0.3 & 190 \\
R4r	 & 10$^{\romn h}$38$^{\romn m}$46\fs200 & +53$^\circ$30\arcmin16\farcs77 & 3.1$\pm$0.3 & 216  & &  R7n	 & 10$^{\romn h}$38$^{\romn m}$45\fs363 & +53$^\circ$30\arcmin17\farcs10 & 3.4$\pm$0.3 & 180 \\
R4s	 & 10$^{\romn h}$38$^{\romn m}$46\fs279 & +53$^\circ$30\arcmin16\farcs72 & 2.7$\pm$0.3 & 212  & &  R7o	 & 10$^{\romn h}$38$^{\romn m}$45\fs549 & +53$^\circ$30\arcmin17\farcs87 & 3.9$\pm$0.5 & 179 \\
R4t	 & 10$^{\romn h}$38$^{\romn m}$46\fs123 & +53$^\circ$30\arcmin16\farcs40 & 2.9$\pm$0.4 & 206  & &  R7p	 & 10$^{\romn h}$38$^{\romn m}$45\fs461 & +53$^\circ$30\arcmin18\farcs84 & 3.2$\pm$0.2 & 177 \\
R4u	 & 10$^{\romn h}$38$^{\romn m}$46\fs350 & +53$^\circ$30\arcmin14\farcs93 & 3.8$\pm$0.4 & 206  & &  R7q	 & 10$^{\romn h}$38$^{\romn m}$45\fs494 & +53$^\circ$30\arcmin16\farcs98 & 3.8$\pm$0.3 & 170 \\
R4v	 & 10$^{\romn h}$38$^{\romn m}$46\fs347 & +53$^\circ$30\arcmin15\farcs14 & 4.0$\pm$0.4 & 205  & &  R7r	 & 10$^{\romn h}$38$^{\romn m}$45\fs533 & +53$^\circ$30\arcmin16\farcs82 & 3.9$\pm$0.3 & 168 \\
R4w	 & 10$^{\romn h}$38$^{\romn m}$46\fs291 & +53$^\circ$30\arcmin16\farcs18 & 2.9$\pm$0.4 & 204  & &  R7s	 & 10$^{\romn h}$38$^{\romn m}$45\fs576 & +53$^\circ$30\arcmin18\farcs10 & 4.0$\pm$0.2 & 165 \\
R4x	 & 10$^{\romn h}$38$^{\romn m}$46\fs308 & +53$^\circ$30\arcmin15\farcs19 & 3.1$\pm$0.3 & 201  & &  R7t	 & 10$^{\romn h}$38$^{\romn m}$45\fs422 & +53$^\circ$30\arcmin18\farcs27 & 2.4$\pm$0.3 & 162 \\
R4y	 & 10$^{\romn h}$38$^{\romn m}$46\fs313 & +53$^\circ$30\arcmin16\farcs20 & 2.9$\pm$0.4 & 194  & &  R7u	 & 10$^{\romn h}$38$^{\romn m}$45\fs306 & +53$^\circ$30\arcmin17\farcs08 & 3.7$\pm$0.4 & 160 \\
R4z	 & 10$^{\romn h}$38$^{\romn m}$46\fs366 & +53$^\circ$30\arcmin14\farcs81 & 3.5$\pm$0.3 & 191  & &  R7v	 & 10$^{\romn h}$38$^{\romn m}$45\fs364 & +53$^\circ$30\arcmin18\farcs99 & 3.7$\pm$0.3 & 157 \\
R4$\alpha$&10$^{\romn h}$38$^{\romn m}$46\fs282 & +53$^\circ$30\arcmin16\farcs48 & 3.3$\pm$0.4 & 187  & &  R7w	 & 10$^{\romn h}$38$^{\romn m}$45\fs423 & +53$^\circ$30\arcmin19\farcs19 & 3.9$\pm$0.2 & 154 \\
R4$\beta$& 10$^{\romn h}$38$^{\romn m}$46\fs296 & +53$^\circ$30\arcmin15\farcs04 & 3.7$\pm$0.3 & 183  & &  R7x	 & 10$^{\romn h}$38$^{\romn m}$45\fs447 & +53$^\circ$30\arcmin18\farcs24 & 3.5$\pm$0.2 & 152 \\
R4$\gamma$&10$^{\romn h}$38$^{\romn m}$46\fs140 & +53$^\circ$30\arcmin16\farcs19 & 3.7$\pm$0.3 & 166  & &  R7y	 & 10$^{\romn h}$38$^{\romn m}$45\fs550 & +53$^\circ$30\arcmin16\farcs96 & 3.9$\pm$0.3 & 147 \\
R4$\delta$&10$^{\romn h}$38$^{\romn m}$46\fs136 & +53$^\circ$30\arcmin15\farcs94 & 2.6$\pm$0.3 & 166  & &  R7z	 & 10$^{\romn h}$38$^{\romn m}$45\fs476 & +53$^\circ$30\arcmin18\farcs34 & 3.3$\pm$0.4 & 145 \\
R5	 & 10$^{\romn h}$38$^{\romn m}$46\fs111 & +53$^\circ$30\arcmin17\farcs03 & 3.4$\pm$0.3 & 606  & &  R7$\alpha$&10$^{\romn h}$38$^{\romn m}$45\fs543 & +53$^\circ$30\arcmin18\farcs70 & 4.7$\pm$0.6 & 143 \\
R5a	 & 10$^{\romn h}$38$^{\romn m}$46\fs119 & +53$^\circ$30\arcmin17\farcs42 & 3.1$\pm$0.4 & 370  & &  R7$\beta$& 10$^{\romn h}$38$^{\romn m}$45\fs545 & +53$^\circ$30\arcmin18\farcs99 & 3.4$\pm$0.4 & 138 \\
R5b	 & 10$^{\romn h}$38$^{\romn m}$46\fs094 & +53$^\circ$30\arcmin17\farcs26 & 3.6$\pm$0.4 & 313  & &  R10	 & 10$^{\romn h}$38$^{\romn m}$45\fs309 & +53$^\circ$30\arcmin08\farcs14 & 3.1$\pm$0.2 & 354 \\
R5c	 & 10$^{\romn h}$38$^{\romn m}$46\fs104 & +53$^\circ$30\arcmin17\farcs79 & 6.2$\pm$0.6 & 262  & &  R10a	 & 10$^{\romn h}$38$^{\romn m}$45\fs335 & +53$^\circ$30\arcmin07\farcs59 & 5.1$\pm$0.5 & 303 \\
R5d	 & 10$^{\romn h}$38$^{\romn m}$46\fs153 & +53$^\circ$30\arcmin17\farcs08 & 3.6$\pm$0.3 & 244  & &  R10b	 & 10$^{\romn h}$38$^{\romn m}$45\fs420 & +53$^\circ$30\arcmin08\farcs06 & 3.8$\pm$0.2 & 270 \\
R5e	 & 10$^{\romn h}$38$^{\romn m}$46\fs071 & +53$^\circ$30\arcmin16\farcs97 & 3.7$\pm$0.2 & 222  & &  R10c	 & 10$^{\romn h}$38$^{\romn m}$45\fs343 & +53$^\circ$30\arcmin07\farcs73 & 3.5$\pm$0.3 & 260 \\
R5f	 & 10$^{\romn h}$38$^{\romn m}$46\fs054 & +53$^\circ$30\arcmin17\farcs48 & 3.0$\pm$0.2 & 206  & &  R10d	 & 10$^{\romn h}$38$^{\romn m}$45\fs338 & +53$^\circ$30\arcmin07\farcs84 & 4.1$\pm$0.4 & 251 \\
R5g	 & 10$^{\romn h}$38$^{\romn m}$46\fs083 & +53$^\circ$30\arcmin16\farcs71 & 3.2$\pm$0.3 & 193  & &  R10e	 & 10$^{\romn h}$38$^{\romn m}$45\fs307 & +53$^\circ$30\arcmin08\farcs28 & 2.8$\pm$0.3 & 235 \\
R6	 & 10$^{\romn h}$38$^{\romn m}$45\fs935 & +53$^\circ$30\arcmin18\farcs20 & 3.4$\pm$0.2 & 378  & &  R10f	 & 10$^{\romn h}$38$^{\romn m}$45\fs298 & +53$^\circ$30\arcmin08\farcs41 & 2.8$\pm$0.4 & 222 \\
R6a	 & 10$^{\romn h}$38$^{\romn m}$46\fs009 & +53$^\circ$30\arcmin17\farcs83 & 2.2$\pm$0.3 & 342  & &  R10g	 & 10$^{\romn h}$38$^{\romn m}$45\fs393 & +53$^\circ$30\arcmin07\farcs84 & 4.4$\pm$0.6 & 200 \\
R6b	 & 10$^{\romn h}$38$^{\romn m}$45\fs879 & +53$^\circ$30\arcmin17\farcs69 & 3.9$\pm$0.2 & 260  & &  R10h	 & 10$^{\romn h}$38$^{\romn m}$45\fs254 & +53$^\circ$30\arcmin07\farcs38 & 4.1$\pm$0.3 & 199 \\
R6c	 & 10$^{\romn h}$38$^{\romn m}$45\fs972 & +53$^\circ$30\arcmin17\farcs51 & 2.7$\pm$0.2 & 258  & &  R10i	 & 10$^{\romn h}$38$^{\romn m}$45\fs395 & +53$^\circ$30\arcmin07\farcs70 & 4.7$\pm$0.5 & 182 \\
R6d	 & 10$^{\romn h}$38$^{\romn m}$45\fs907 & +53$^\circ$30\arcmin17\farcs88 & 4.2$\pm$0.3 & 242  & &  R10j	 & 10$^{\romn h}$38$^{\romn m}$45\fs371 & +53$^\circ$30\arcmin08\farcs15 & 4.0$\pm$0.5 & 177 \\
R6e	 & 10$^{\romn h}$38$^{\romn m}$46\fs005 & +53$^\circ$30\arcmin17\farcs46 & 3.9$\pm$0.4 & 221  & &  R10k	 & 10$^{\romn h}$38$^{\romn m}$45\fs421 & +53$^\circ$30\arcmin08\farcs27 & 3.8$\pm$0.3 & 177 \\
R6f	 & 10$^{\romn h}$38$^{\romn m}$45\fs952 & +53$^\circ$30\arcmin18\farcs44 & 3.7$\pm$0.2 & 215  & &  R10l	 & 10$^{\romn h}$38$^{\romn m}$45\fs408 & +53$^\circ$30\arcmin07\farcs80 & 3.9$\pm$0.3 & 173 \\
R6g	 & 10$^{\romn h}$38$^{\romn m}$45\fs907 & +53$^\circ$30\arcmin16\farcs99 & 3.4$\pm$0.4 & 170  & &  R10m	 & 10$^{\romn h}$38$^{\romn m}$45\fs374 & +53$^\circ$30\arcmin07\farcs86 & 3.1$\pm$0.3 & 169 \\
R6h	 & 10$^{\romn h}$38$^{\romn m}$45\fs948 & +53$^\circ$30\arcmin17\farcs94 & 3.1$\pm$0.3 & 168  & &  R10n	 & 10$^{\romn h}$38$^{\romn m}$45\fs424 & +53$^\circ$30\arcmin07\farcs90 & 3.1$\pm$0.4 & 168 \\
R6i	 & 10$^{\romn h}$38$^{\romn m}$45\fs965 & +53$^\circ$30\arcmin17\farcs75 & 2.5$\pm$0.2 & 168  & &  N	 & 10$^{\romn h}$38$^{\romn m}$45\fs893 & +53$^\circ$30\arcmin11\farcs47 &14.2$\pm$1.0 & 3735\\
R6j	 & 10$^{\romn h}$38$^{\romn m}$46\fs007 & +53$^\circ$30\arcmin16\farcs94 & 4.0$\pm$0.5 & 165  & &  Na	 & 10$^{\romn h}$38$^{\romn m}$45\fs970 & +53$^\circ$30\arcmin10\farcs85 & 5.8$\pm$0.6 & 370 \\
R6k	 & 10$^{\romn h}$38$^{\romn m}$45\fs988 & +53$^\circ$30\arcmin17\farcs28 & 3.1$\pm$0.2 & 160  \\
R6l	 & 10$^{\romn h}$38$^{\romn m}$46\fs001 & +53$^\circ$30\arcmin17\farcs19 & 3.7$\pm$0.4 & 156  \\
R6m	 & 10$^{\romn h}$38$^{\romn m}$45\fs967 & +53$^\circ$30\arcmin17\farcs12 & 4.2$\pm$0.5 & 146  \\

\cline{1-5}\cline{7-11} 
\end{tabular}
\label{knots}
\end{table*}


\label{masses}


Upper limits to the dynamical masses (M$_{\ast}$) inside the half
light radius (R) for each observed knot have been estimated under the
following assumptions: (i) the systems are spherically symmetric; (ii) they 
are gravitationally bound; and (iii) they have isotropic velocity
distributions [$\sigma^2$(total)\,=\,3 $\sigma_{\ast}^2$]. 
The general expression for the virial mass of a cluster is
$\eta$\,$\sigma_{\ast}^2$\,R/G, where R is the effective gravitational
radius and $\eta$ is a dimensionless number that takes into account departures 
from isotropy in the velocity distribution and the spatial mass distribution, binary
fraction, mean surface density,
etc.\ \citep{2005ApJ...620L..27B,2006MNRAS.369.1392F}. Following
\cite{1996ApJ...466L..83H,1996ApJ...472..600H}, and for consistence with
Paper~I, Paper~II
and \cite{2008PhDT........35H}, we obtain the dynamical masses inside the
half-light radius 
using $\eta$\,=\,3 and adopting the half-light radius as a reasonable
approximation of the effective radius. Other authors
\citep[e.g.][]{1987degc.book.....S,2001MNRAS.326.1027S,2008A&A...490..125M}
assumed that the $\eta$ value is about 9.75 to obtain the total mass. 
On the absence of any knowledge about the tidal radius of the clusters, we
adopted this conservative approach. On the derived
masses, the different adopted $\eta$ values act as multiplicative factors.

It must be noted that while we can measure the size of each knot, we
do not have direct access  
to the stellar velocity dispersion of each individual cluster, since
our spectroscopic measurements encompass a wider area
(1.0\,$\times$\,1.9\,arcsec$^2$ which corresponds approximately to
73\,$\times$\,131\,pc$^2$ at the adopted distance for NGC\,3310) that includes
the whole 
CNSFRs to which each group of knots belongs. The use of these wider size scale
velocity dispersion measurements to estimate the mass of each knot, leads us to
overestimate the mass of the individual clusters, and hence of each CNSFR.
As we can not be sure that we are actually measuring their velocity dispersion,
we prefer to say that our measurements of $\sigma_{\ast}$, and hence the
dynamical masses, constitute upper limits. Although we are well aware of the
difficulties, still we are confident that these upper limits are valid and
important for comparison with the gas kinematical measurements. 

The estimated dynamical masses for each knot and their corresponding errors 
are listed in Table \ref{mass}. For the regions that have been observed in
more than one slit position, we list the derived values using the two
separate stellar velocity dispersions. The dynamical masses in the rows
labelled  ``sum'' have been found by adding the individual masses in a given
CNSFR, as well as the galaxy nucleus, N. The fractional errors of the
dynamical masses of the individual knots and of the CNSFRs are listed in
column 4.
As explained above, since the stellar velocity dispersion for
the R1+R2 region of NGC\,3310 has a large error,
we do not list the errors of the dynamical masses of the individual knots or
of the whole region R12sum.


\begin{table*}
\centering
\caption[]{{\bf Upper limits to the} dynamical masses.}
\begin{tabular} {@{}lc c c @{}p{0.5cm}@{} lc c c@{}p{0.5cm}@{} lc c c @{}}
\cline{1-4}\cline{6-9}\cline{11-14}
 Region & Slit & M$_{\ast}$ & error(\%)  & & Region & Slit & M$_{\ast}$ & error(\%)   & & Region & Slit & M$_{\ast}$ & error(\%)\\ 
  
\cline{1-4}\cline{6-9}\cline{11-14}

R1	    & S2 & 202:          & ---&  & R4n	    & S2 & 35$\pm$7	 & 19 &  & R7	  & S2 & 42$\pm$9	 & 22\\
R1a	    & S2 & 115:          & ---&  & R4o	    & S2 & 32$\pm$6	 & 18 &  & R7a    & S2 & 71$\pm$15	 & 21\\
R2	    & S2 & 208:          & ---&  & R4p	    & S2 & 25$\pm$5	 & 19 &  & R7b    & S2 & 46$\pm$11	 & 23\\
R2a	    & S2 & 241:          & ---&  & R4q	    & S2 & 32$\pm$7	 & 21 &  & R7c    & S2 & 64$\pm$14	 & 21\\
R2b	    & S2 & 144:          & ---&  & R4r	    & S2 & 31$\pm$6	 & 19 &  & R7d    & S2 & 40$\pm$9	 & 22\\  
R12sum	    & S2 & 911:          & ---&  & R4s	    & S2 & 27$\pm$6	 & 20 &  & R7e    & S2 & 32$\pm$8	 & 24\\  
R4	    & S1 & 26$\pm$4	 & 17 &  & R4t	    & S2 & 30$\pm$6	 & 22 &  & R7f    & S2 & 52$\pm$11	 & 22\\  
R4a	    & S1 & 27$\pm$5	 & 17 &  & R4u	    & S2 & 39$\pm$8	 & 20 &  & R7g    & S2 & 60$\pm$13	 & 22\\  
R4b	    & S1 & 22$\pm$4	 & 18 &  & R4v	    & S2 & 41$\pm$8	 & 19 &  & R7h    & S2 & 51$\pm$11	 & 22\\	  
R4c	    & S1 & 21$\pm$4	 & 18 &  & R4w	    & S2 & 30$\pm$6	 & 22 &  & R7i    & S2 & 40$\pm$9	 & 22\\  
R4d	    & S1 & 36$\pm$6	 & 16 &  & R4x	    & S2 & 32$\pm$6	 & 19 &  & R7j    & S2 & 53$\pm$13	 & 24\\	  
R4e	    & S1 & 38$\pm$6	 & 17 &  & R4y	    & S2 & 30$\pm$6	 & 22 &  & R7k    & S2 & 44$\pm$10	 & 23\\	  
R4f	    & S1 & 38$\pm$6	 & 16 &  & R4z	    & S2 & 36$\pm$7	 & 19 &  & R7l    & S2 & 54$\pm$12	 & 22\\	  
R4g	    & S1 & 26$\pm$4	 & 17 &  & R4$\alpha$  & S2 & 33$\pm$7	 & 21 &  & R7m    & S2 & 40$\pm$9	 & 23\\	  
R4h	    & S1 & 25$\pm$5	 & 21 &  & R4$\beta$   & S2 & 38$\pm$7	 & 19 &  & R7n    & S2 & 44$\pm$10	 & 23\\	  
R4i	    & S1 & 37$\pm$6	 & 16 &  & R4$\gamma$  & S2 & 38$\pm$7	 & 19 &  & R7o    & S2 & 52$\pm$13	 & 25\\	  
R4j	    & S1 & 27$\pm$5	 & 19 &  & R4$\delta$  & S2 & 27$\pm$5	 & 20 &  & R7p    & S2 & 43$\pm$9	 & 22\\	  
R4k	    & S1 & 24$\pm$4	 & 17 &  & R5	    & S2 & 35$\pm$7	 & 19 &  & R7q    & S2 & 50$\pm$11	 & 22\\	  
R4l	    & S1 & 30$\pm$7	 & 21 &  & R5a	    & S2 & 32$\pm$7	 & 21 &  & R7r    & S2 & 52$\pm$12	 & 22\\	  
R4m	    & S1 & 25$\pm$5	 & 19 &  & R5b	    & S2 & 36$\pm$7	 & 20 &  & R7s    & S2 & 53$\pm$12	 & 22\\  
R4n	    & S1 & 30$\pm$5	 & 18 &  & R5c	    & S2 & 63$\pm$12	 & 19 &  & R7t    & S2 & 32$\pm$8	 & 24\\	  
R4o	    & S1 & 28$\pm$5	 & 17 &  & R5d	    & S2 & 36$\pm$7	 & 19 &  & R7u    & S2 & 48$\pm$11	 & 24\\	  
R4p	    & S1 & 21$\pm$4	 & 18 &  & R5e	    & S2 & 37$\pm$7	 & 18 &  & R7v    & S2 & 48$\pm$11	 & 23\\	  
R4q	    & S1 & 28$\pm$6	 & 20 &  & R5f	    & S2 & 30$\pm$5	 & 18 &  & R7w    & S2 & 52$\pm$11	 & 22\\	  
R4r	    & S1 & 27$\pm$5	 & 18 &  & R5g	    & S2 & 33$\pm$6	 & 19 &  & R7x    & S2 & 46$\pm$10	 & 22\\	  
R4s	    & S1 & 24$\pm$5	 & 19 &  & R45sum   & S2 & 1317$\pm$41   & 3  &  & R7y    & S2 & 52$\pm$12	 & 22\\	  
R4t	    & S1 & 26$\pm$5	 & 21 &  & R6	    & S2 & 28$\pm$8	 & 30 &  & R7z    & S2 & 44$\pm$11	 & 24\\	  
R4u	    & S1 & 34$\pm$6	 & 19 &  & R6a	    & S2 & 18$\pm$6	 & 32 &  & R7$\alpha$  & S2 & 63$\pm$15	 & 25\\	  
R4v	    & S1 & 36$\pm$7	 & 19 &  & R6b	    & S2 & 32$\pm$9	 & 29 &  & R7$\beta$   & S2 & 44$\pm$11	 & 24\\	  
R4w	    & S1 & 26$\pm$5	 & 21 &  & R6c	    & S2 & 22$\pm$7	 & 30 &  & R7sum  & S2 & 1413$\pm$60	 & 4 \\	  
R4x	    & S1 & 28$\pm$5	 & 18 &  & R6d	    & S2 & 35$\pm$10	 & 30 &  & R10    & S1 & 33$\pm$5	 & 17\\	  
R4y	    & S1 & 26$\pm$5	 & 21 &  & R6e	    & S2 & 32$\pm$10	 & 31 &  & R10a   & S1 & 53$\pm$10	 & 18\\	  
R4z	    & S1 & 31$\pm$6	 & 18 &  & R6f	    & S2 & 30$\pm$9	 & 29 &  & R10b   & S1 & 40$\pm$6	 & 16\\	  
R4$\alpha$  & S1 & 29$\pm$6	 & 20 &  & R6g	    & S2 & 28$\pm$9	 & 31 &  & R10c   & S1 & 37$\pm$6	 & 18\\	  
R4$\beta$   & S1 & 33$\pm$6	 & 18 &  & R6h	    & S2 & 26$\pm$8	 & 30 &  & R10d   & S1 & 43$\pm$8	 & 18\\	  
R4$\gamma$  & S1 & 33$\pm$6	 & 18 &  & R6i	    & S2 & 21$\pm$6	 & 30 &  & R10e   & S1 & 29$\pm$5	 & 19\\	  
R4$\delta$  & S1 & 23$\pm$5	 & 19 &  & R6j	    & S2 & 33$\pm$10	 & 31 &  & R10f   & S1 & 29$\pm$6	 & 21\\	  
R4sum	    & S1 & 886$\pm$30	 & 3  &  & R6k	    & S2 & 25$\pm$8	 & 30 &  & R10g   & S1 & 46$\pm$9	 & 21\\	  
R4	    & S2 & 30$\pm$5	 & 18 &  & R6l	    & S2 & 31$\pm$10	 & 31 &  & R10h   & S1 & 43$\pm$7	 & 17\\	  
R4a	    & S2 & 31$\pm$6	 & 18 &  & R6m	    & S2 & 35$\pm$11	 & 31 &  & R10i   & S1 & 50$\pm$9	 & 19\\	  
R4b	    & S2 & 25$\pm$5	 & 19 &  & R6sum    & S2 & 397$\pm$33    & 8  &  & R10j   & S1 & 42$\pm$8	 & 19\\	  
R4c	    & S2 & 25$\pm$5	 & 19 &  & S6	    & S2 & 28$\pm$6	 & 23 &  & R10k   & S1 & 40$\pm$7	 & 17\\	  
R4d	    & S2 & 42$\pm$7	 & 17 &  & S6a	    & S2 & 37$\pm$8	 & 23 &  & R10l   & S1 & 40$\pm$7	 & 17\\	  
R4e	    & S2 & 43$\pm$8	 & 18 &  & S6b	    & S2 & 18$\pm$5	 & 25 &  & R10m   & S1 & 32$\pm$6	 & 18\\	  
R4f	    & S2 & 43$\pm$7	 & 17 &  & S6c	    & S2 & 23$\pm$6	 & 24 &  & R10n   & S1 & 33$\pm$7	 & 20\\	  
R4g	    & S2 & 30$\pm$5	 & 18 &  & S6d	    & S2 & 18$\pm$5	 & 25 &  & R10sum & S1 & 588$\pm$28	 & 5 \\	  
R4h	    & S2 & 28$\pm$6	 & 22 &  & S6e	    & S2 & 20$\pm$5	 & 23 &  & N      & S1 & 526$\pm$57	 & 11\\	  
R4i	    & S2 & 42$\pm$7	 & 17 &  & S6f	    & S2 & 24$\pm$6	 & 23 &  & Na     & S1 & 216$\pm$28	 & 13\\	  
R4j	    & S2 & 30$\pm$6	 & 19 &  & S6g	    & S2 & 20$\pm$5	 & 25 &  & Nsum   & S1 & 742$\pm$64	 & 9 \\	  
R4k	    & S2 & 27$\pm$5	 & 18 &  & S6h	    & S2 & 19$\pm$5	 & 25 &  & \\	  
R4l	    & S2 & 35$\pm$8	 & 22 &  & S6sum    & S2 & 210$\pm$17    & 8  &  & \\	  
R4m	    & S2 & 29$\pm$6	 & 20 & &  & & \\

\cline{1-4}\cline{6-9}\cline{11-14}
\multicolumn{9}{l}{masses in 10$^5$ M$_\odot$.}
\end{tabular}
\label{mass}
\end{table*}


\section{Ionizing star cluster properties}

For each of the CNSFR the total number of ionizing photons
was derived from the total observed H$\alpha$ luminosities given by
\cite{2000MNRAS.311..120D} and \cite{1993MNRAS.260..177P},
correcting for the different assumed distance. \cite{2000MNRAS.311..120D}
and \cite{1993MNRAS.260..177P} give the already extinction corrected luminosities. 
\cite{2000MNRAS.311..120D} estimated a diameter of
2\,arcsec for regions R2, R4, R6, R7 and R10, 1.4\,arcsec for R1 and R5,
and 3.4 for the Jumbo region. For this latter region we give the values  
derived using the different  quantities for the luminosities given by
\cite[][region R19]{2000MNRAS.311..120D} and \cite[][region
A]{1993MNRAS.260..177P}. For R1+R2 and R4+R5 (S2) we added their
H$\alpha$ luminosities. No values are found in the literature for the
H$\alpha$ luminosity of regions X and Y. Our derived values of Q(H$_0$)
constitute lower limits 
since we have not taken into account the fraction of photons that may have
been absorbed by dust or may have escaped the region.   

Once the number of Lyman continuum photons have been calculated, the masses of the
ionizing star clusters, M$_{ion}$, have been derived using the solar
metallicity single burst models by \cite{1995A&AS..112...35G} which provide
the number of ionizing photons per unit mass, [$Q(H_0)/M_{ion}$]. 
A Salpeter initial mass function \citep[][IMF]{1955ApJ...121..161S} has been
assumed with lower and upper mass limits of 0.8 and 120 M$_\odot$. In order to
take into account the evolution of the \HII\ region, we have made use of the
fact that a relation exists between the degree of evolution of the cluster, as
represented by the equivalent width of the H$\beta$ emission line, and the
number of Lyman continuum photons per unit solar mass \citep[e.g.\
][]{2000MNRAS.318..462D}. We have  measured the EW(H$\beta$) from our spectra
(see Table  \ref{parameters}) following the same procedure as in
\cite{2006MNRAS.372..293H,2008MNRAS.383..209H}, that is defining a
pseudo-continuum to take into account the absorption  from the underlying
stellar population. This procedure in fact may underestimate the value of the
equivalent width, since it includes the contribution to the continuum by the
older stellar population  (see discussions in \citealt{2007MNRAS.382..251D}
and \citealt{2008A&A...482...59D}). The derived 
masses for the ionizing populations of the observed CNSFRs are given in column
6 of Table \ref{parameters} and are between 1 and 7 per cent of the dynamical
mass (see column 9 of the table).


\begin{table*}
\centering
\caption[]{Physical parameters.}
\begin{tabular} {@{} l c c c c c c c c @{} }
\hline
 Region & c(H$\alpha$) &L(H$\alpha$) & Q(H$_0$)  & 
 EW(H$\beta$) & M$_{ion}$ & N$_e^c$ & M$_{{\rm HII}}$ &
 M$_{ion}$/M$_{\ast}$   \\ 
  &  &  &  &  &  &  &  &  (per cent)  \\
\hline

R1+R2 & 0.23$^a$ & 102.0$^a$ & 74.9 & 28.6 & 13.9 & 100$^c$ & 3.38 & 1.5 \\
R4    & 0.23$^a$ & 144.0$^a$ &106.0 & 32.4 & 17.6 & 100$^b$ & 4.78 & 1.6 \\ 
R4+R5 & 0.20$^a$ & 218.0$^a$ &160.0 & 41.7 & 21.4 & 100$^b$ & 7.24 & 1.1 \\
R6    & 0.17$^a$ &  57.3$^a$ & 42.1 & 16.7 & 12.4 & 100$^c$ & 1.90 & 3.5 \\
S6    & 0.35$^b$ &  62.5$^b$ & 45.9 & 12.5 & 17.4 & 100$^b$ & 2.07 & 6.6 \\
R7    & 0.17$^a$ &  45.5$^a$ & 33.5 & 19.4 &  8.7 & 100$^c$ & 1.51 & 1.0 \\
R10   & 0.23$^a$ &  45.5$^a$ & 33.5 &  9.7 & 15.7 & 100$^c$ & 1.51 & 2.4 \\
N     & 0.42$^b$ & 113.0$^b$ & 82.9 & 11.0 & 34.9 & 800$^d$ & 0.47 & 4.7\\[2pt]
J     & 0.23$^a$ & 573.0$^a$ &421.0 & 82.5 & 31.4 & 200$^b$ & 9.52 & --- \\
      & 0.19$^b$ & 236.0$^b$ &174.0 & 82.5 & 12.9 & 200$^b$ & 3.93 & --- \\

\hline
\multicolumn{9}{@{}p{0.65\textwidth}}{Note. Luminosities in
  10$^{38}$\,erg\,s$^{-1}$, masses in 10$^5$ M$_\odot$, ionizing photons in
  10$^{50}$\,photon\,s$^{-1}$ and densities in cm$^{-3}$}\\ 
\multicolumn{9}{@{}l}{$^a$From \citet{2000MNRAS.311..120D}.} \\
\multicolumn{9}{@{}l}{$^b$From \citet{1993MNRAS.260..177P}.} \\
\multicolumn{9}{@{}p{0.65\textwidth}}{$^c$Assumed from the values given by
  \cite{1993MNRAS.260..177P} for the other CNSFRs of this galaxy studied by
  them.}\\
\multicolumn{9}{@{}p{0.65\textwidth}}{$^d$Derived from the spectrum of the
  nucleus of NGC\,3310 plotted in Fig.\ 5a of
  \cite{1993MNRAS.260..177P} (see discussion in the text).} 
\end{tabular}
\label{parameters}
\end{table*}

\begin{figure}
\centering
\vspace*{0.3cm}
\includegraphics[width=.44\textwidth,angle=0]{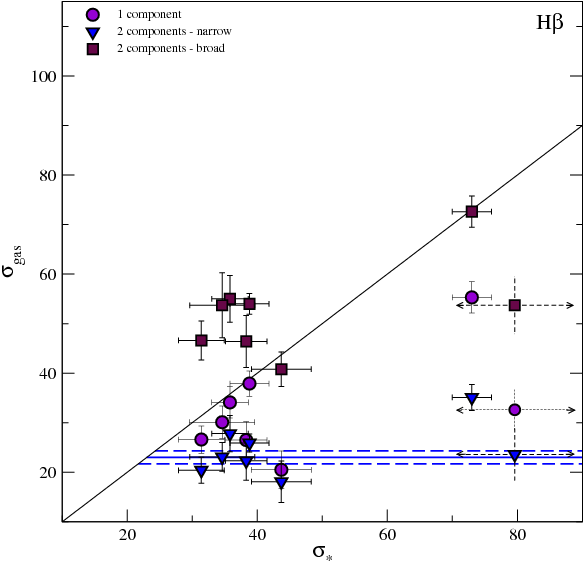}\\\vspace*{0.5cm}
\includegraphics[width=.44\textwidth,angle=0]{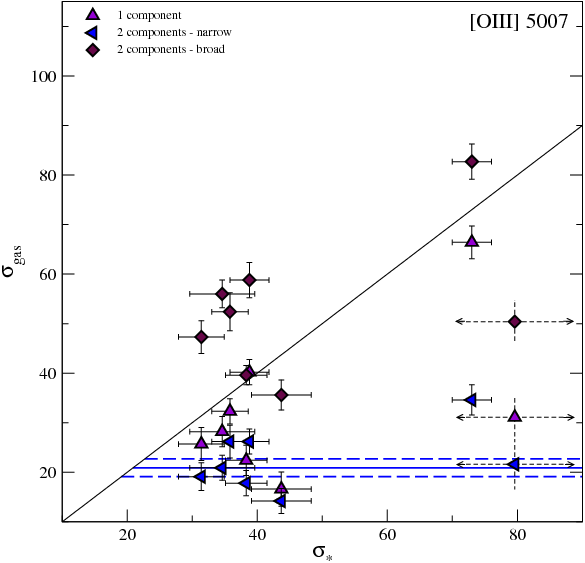}
\caption[]{Upper panel:  relation between velocity dispersions of the gas
  (derived from H$\beta$) and stars (CaT) for the CNSFRs and the nucleus of
  NGC\,3310. Symbols are as follows: Upper panel:
  single Gaussian fit, violet circles; two Gaussian fit, broad component,
  maroon squares; narrow component, blue downward triangles; lower
  panel:  as the upper panel for the [O{\sc iii}] line. Violet upward
  triangles correspond to the estimates using a single 
  Gaussian fit, maroon diamonds represent the broad components of the two 
  Gaussian fit and blue left triangles, the narrow components. The
  blue solid line represents the average velocity dispersion of the
  narrow component of the gas (H$\beta$ upper and [O{\sc iii}] lower panel),
  for the CNSFRs, and the blue dashed lines represent their estimated
  errors.[{\it
  See the electronic edition of the Journal for a   
  colour version of this figure.}]} 
\label{dispersions3310}
\end{figure}

The amount of ionized gas (M$_{{\rm HII}}$) associated to each star-forming
region complex can also be derived from the H$\alpha$ luminosities 
using the electron density (N$_e$) dependency relation given by
\cite{1990ApJ...356..389M} for an electron temperature of 10$^4$\,K.
The electron density for each
region (obtained from  the [S{\sc ii}]\,$\lambda\lambda$\,6717\,/\,6731\,\AA\
line ratio) has been taken from \cite{1993MNRAS.260..177P} for the regions in
common. For the rest of the regions we assume a value of N$_e$ equal to
100\,cm$^{-3}$, the value given by \cite{1993MNRAS.260..177P} for the other
CNSFRs of this galaxy studied by them. 
For the nucleus of NGC\,3310 we find an anomalous behaviour when we estimate
M$_{{\rm HII}}$ using the values of the density (8200\,cm$^{-3}$) given by
\cite{1993MNRAS.260..177P}. We derive a rather low value of M$_{{\rm HII}}$
(5\,$\times$\,10$^3$\,\Msol) for the H$\alpha$ luminosity calculated by these
authors. Then, when reviewing the density
reported by Pastoriza and collaborators we found that it is consistent with
the ratio of about 0.5 between the 
intensity values of the [S{\sc ii}]\,$\lambda\lambda$\,6717,6731\,\AA\
emission lines given by them. However, a careful inspection of the spectrum
of the nucleus of NGC\,3310 plotted in Fig.\ 5a of
\cite{1993MNRAS.260..177P} reveals that the intensities of these two [\SII]
emission lines are very similar to each other (ratio of about 0.9). A rough
estimate using this revised ratio yields an electron density value of
about 800\,cm$^{-3}$. The corresponding M$_{{\rm HII}}$ value derived using
this lower density is an order of magnitude higher.

\section{Discussion}

\label{Velocity dispersions}

\subsection{Star and gas kinematics}

Fig.\ \ref{dispersions3310} shows the relations between the velocity
dispersions of gas and stars for NGC\,3310. In the upper and lower panels we
plot the relation derived from the H$\beta$ and [\OIII] emission lines,
respectively. In both panels the straight line shows the one-to-one 
relation. In the upper panel the violet\footnote{In all figures, colours can be
seen in the electronic version of the paper.} circles show the gas velocity
dispersion measured from the H$\beta$ emission line using a single Gaussian
fit, maroon squares and blue downward triangles show the values measured from
the broad and narrow components respectively using two-component Gaussian
fits. The deviant points, marked with arrows, correspond to region 
R1+R2, which has a spectrum with low signal-to-noise ratio for which the
cross-correlation method 
does not provide accurate results. In the lower panel,
violet upward triangles, maroon diamonds and blue left triangles correspond to
the values obtained by a single Gaussian fit, and to the broad and narrow
components of the two Gaussian fits, respectively. Again, the deviant points
(R1+R2) are marked with arrows.

The H$\beta$ velocity dispersions of the CNSFRs of NGC\,3310 derived by a
single-component Gaussian fit are the same, within the errors, as the 
stellar ones, except for R1+R2 and R7, for which they are lower by about 50
and 25\,km\,s$^{-1}$, respectively. For R1+R2 this can be due to an
overestimate of $\sigma_\ast$. Given the relatively low metal 
abundance of NGC\,3310 \citep{1993MNRAS.260..177P}  the emission lines
have generally a very good signal-to-noise ratio in the spectra of the CNSFRs, while for the
red continuum it depends on the particular case (see Figs.\ \ref{profiles3310}
and \ref{spectra3310} for the spatial profiles and the spectra,
respectively). On the other hand, the stellar velocity dispersions of the
CNSFRs are lower than those of the broad component of H$\beta$ by about
20\,km\,s$^{-1}$, again except for R1+R2 where $\sigma_\ast$ is greater by
about 35\,km\,s$^{-1}$, and for R7 where $\sigma_\ast$ and H$\beta$ broad are
in good agreement. The narrow component of the CNSFRs shows velocity
dispersions very similar for all the regions, with an average value of
23.0\,$\pm$\,1.3\,km\,s$^{-1}$, and is represented as a blue solid line 
in the upper panel of Fig.\ \ref{dispersions3310}, while the dashed lines of
the same colour represent its error. 

The [O{\sc iii}] emission lines show the same behaviour as the H$\beta$
lines for the two different components. In all cases the gas narrow component
has velocity dispersion 
lower than the stellar one, and the broad component is slightly over it.
The broad component of the
[O{\sc iii}]\,5007\,\AA\ emission line (maroon diamonds in the figure) is
wider than the stellar lines by about 20\,km\,s$^{-1}$, except for R4+R5 (S2
slit) and R7 for which they are approximately similar. The narrow component of
[O{\sc iii}] (blue left triangles in the 
figure) shows a relatively constant value with an average of
20.9\,$\pm$\,1.8\,km\,s$^{-1}$ (blue lines in the figure). 

In general, the broad and narrow components of the H$\beta$ line have comparable
fluxes. The ratios between the fluxes in the narrow and broad
components of the H$\beta$ emission line of the regions (including J, X and
Y) vary from 0.95 to 1.65, except for R10 and R7 for which we find ratios of
0.65 and 2.5, respectively. 
The ratio of the narrow to broad [O{\sc iii}] fluxes is between 0.85 and 1.5;
R10 and R7, for which this value is 0.65 and 1.98 respectively, are again the
exception. 

The behaviour of the velocity dispersions of the CNSFRs is
found to be different in NGC\,3310 from what was found for NGC\,2903 and
NGC\,3351 (Papers I and II). For NGC\,3310 we find a very good agreement in most cases
between velocity dispersions derived from gaseous and stellar lines assuming
a single component for the gas. For NGC\,2903 and 
NGC\,3351, in contrast, the velocity dispersions derived from emission lines
measured using a single Gaussian fit are very different from the stellar ones
(lower for H$\beta$ and  higher for [O{\sc iii}]) by
about 20\,km\,s$^{-1}$; see discussion in Paper II). Remarkably, the velocity
dispersions derived from the narrow components of the two Gaussian fits show a
relatively constant value for the whole CNSFR sample (about 23\,kms$^{-1}$).
If this narrow component is identified with ionized
gas in a rotating disc, therefore supported by rotation, then the broad
component could, in principle, correspond to the gas response to the
gravitational potential of the stellar cluster, supported by dynamical
pressure, explaining the coincidence with the stellar velocity dispersion in
most cases in the H$\beta$ line \citep[see ][and references 
  therein]{2004A&A...424..447P}. The velocity excess shown by the broad
component of H$\beta$ and [O{\sc iii}] in most CNSFRs of NGC\,3310 
could be identified with peculiar velocities in the ionized gas (mainly
in the high ionization gas) related to massive star
winds or even supernova remnants.

Our interpretation of the emission line structures, in the present study and
in Papers~I and II, parallels that of the studies of Westmoquette and
collaborators 
(see for example Westmoquette et al.\ 2007a,b). They observed a narrow
($\sim$\,35-100\,km\,s$^{-1}$) and a broad ($\sim$\,100-400\,km\,s$^{-1}$)
component to the H$\alpha$ line across their four fields in the dwarf
galaxy NGC\,1569. They conclude that ``the most likely explanation of the narrow
component is that it represents the general disturbed optically emitting
ionized interstellar medium (ISM), arising through a convolution of the
stirring effects of the starburst and gravitational virial motions''. They also
conclude that ``the broad component results from the highly turbulent velocity
field associated with the interaction of the hot phase of the ISM
(material that is photo-evaporated or thermally evaporated through the action of
the strong ambient radiation field, or mechanically ablated by the impact of
fast-flowing cluster winds) with cooler gas knots, setting up turbulent mixing 
layers \citep[e.g.][]{1990MNRAS.244P..26B,1993ApJ...407...83S}''.
However, our broad component velocity dispersion values derived from H$\beta$,
all significantly lower than 100\,km\,s$^{-1}$ and similar  to the stellar
velocity dispersions,  resemble more their narrow component values. 
This is the same behaviour that we find in the broad component of the [\OIII]
emission line for the CNSFRs of NGC\,3310. 
The  [O{\sc iii}] line profiles of the regions in Papers~I and II on the other
hand, behave as those described  by Westmoquette and collaborators. 

\nocite{2007MNRAS.381..894W,2007MNRAS.381..913W}
 
There are other studies that identified an underlying broad
component to the recombination emission lines 
such as \cite{1987MNRAS.226...19D,1996MNRAS.279.1219T} in the M33 giant \HII\
region NGC\,604;
\cite*{1994ApJ...425..720C,1999MNRAS.302..677M} in the central region
of 30 Doradus; \cite{1997ApJ...488..652M} in four Wolf-Rayet galaxies, and
\cite{1999ApJ...522..199H} in the starburst galaxy NGC\,7673.
More recently, Westmoquette et al.~(2007c, 2009)
\nocite{2007ApJ...671..358W,2009ApJ...696..192W} found a
broad feature in the H$\alpha$ emission lines in the starburst core of M82;
\cite{2007A&A...461..471O} and \cite{2009arXiv0903.2280J} in the blue compact
galaxies ESO\,338-IG04 and Mrk\,996, respectively;
\cite{2006MNRAS.370..799S,firposubmitted} in giant extragalactic \HII\ regions, and
\cite{2008PhDT........35H} in CNSFRs of early type
spiral galaxies. The first two and the last three studies also found this broad
component in the forbidden emission lines.

In the nuclear region of the galaxy the stars and the broad component of the
ionized gas, both for H$\beta$ and [O{\sc iii}] have the same width, with a
$\sigma$ of about 73 and 83\,km\,s$^{-1}$, respectively. The profiles of
the nuclear lines hence, behave similarly to what is  described by
Westmoquette and collaborators. The gas narrow component however shows a
substantially lower velocity dispersion though still higher by about
15\,km\,s$^{-1}$, than that of the CNSFRs. 

For regions J, X and Y we
can not derive a stellar velocity dispersion due to the low signal-to-noise
ratio of their continuum and the noise added by subtracting the large amount
of emission lines present in their red spectra
(see right panels of Fig.\ \ref{spectra3310}). However, if we analyze the
velocity dispersions derived from their strong emission lines in the blue
spectral range we find values very similar to those of the other
CNSFRs (see Table \ref{disp}).


%

\label{Radial velocities 3310}

\begin{figure}
\centering
\includegraphics[width=.46\textwidth,angle=0]{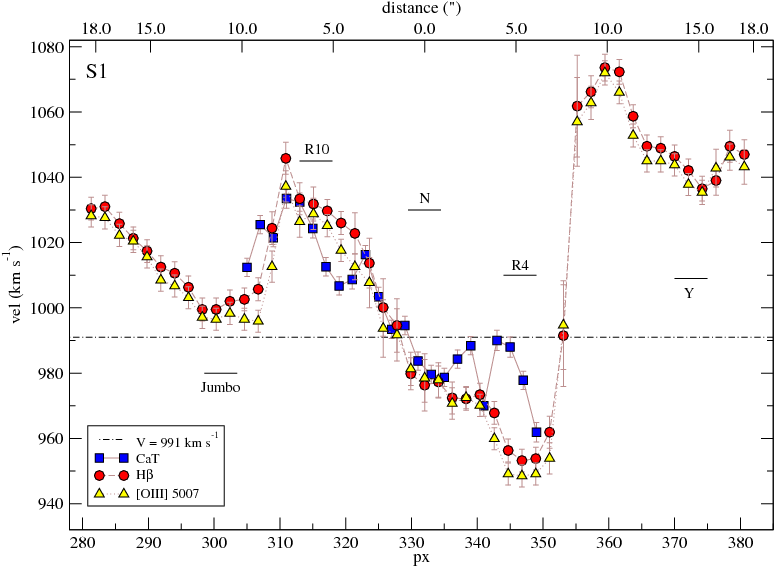}\\
\vspace*{0.3cm}
\includegraphics[width=.46\textwidth,angle=0]{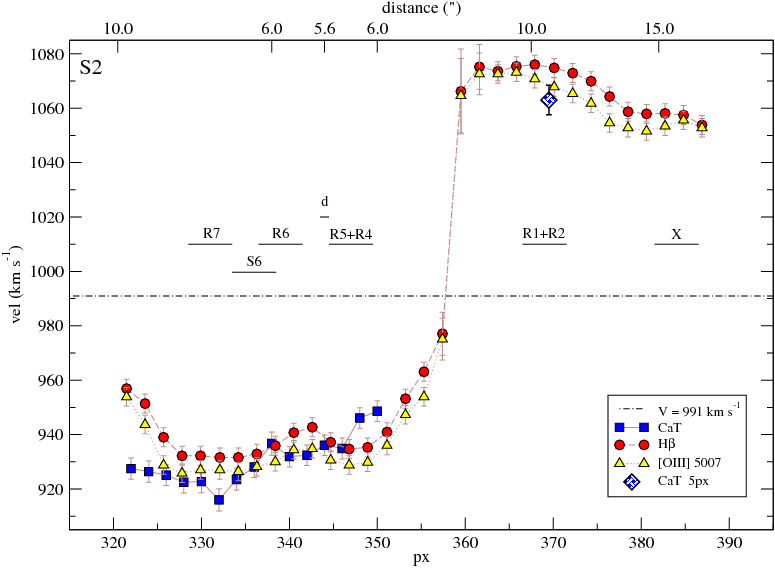}
\caption[]{Radial velocities along the slit versus pixel number for each slit
  position of NGC\,3310 (upper panel: S1; lower panel: S2) as
  derived from the gas emission lines (red circles: H$\beta$; upward triangles:
  [O{\sc iii}]) and the stellar absorption ones (blue squares). The
  criss-crossed white-blue diamond in the lower panel correspond to the
  stellar velocity of R1+R2 
  derived using the 5\,px aperture. The individual
  CNSFRs and the nucleus, ``N'', or the closest position to it, ``d'', are
  marked in the plots. The dashed-dotted line is the
  velocity of the nucleus of NGC\,3310 derived by
  \citet{1976A&A....49..161V}. The distance in arcsec from
  the nucleus is displayed in the upper x-axis of each panel. [{\it
  See the electronic edition of the Journal for a   
  colour version of this figure.}]}
\label{velocities-3310}
\end{figure}

Fig.\ \ref{velocities-3310} presents the radial velocities derived from the 
H$\beta$ and [O{\sc iii}] emission lines and the CaT absorptions along the
slit for each angular slit position of NGC\,3310, S1 in the upper panel and S2
in the lower one. The rotation curves seem to have the turnover points located
in or near the star-forming ring. 
Due to the relatively low
metallicity of NGC\,3310 \citep{1993MNRAS.260..177P}, the gas emission lines
are very strong in the central zone of the galaxy and we can derive the
radial velocity of the gas much further (up to 18\,\arcsec) than for the stellar CaT absorptions.  
The gas H$\beta$ and [O{\sc iii}]\,5007\,\AA\ radial velocities and the
stellar ones are in very good agreement, except in the zone located around R4,
where we find a difference between gas and stars of about 30 to
40\,km\,s$^{-1}$. These differences could be due to a
low signal-to-noise ratio of the stellar continuum emission in this zone of
the galaxy. 

In both slit positions we can appreciate a very steep slope in the radial
velocity curve as measured from  H$\beta$ and [O{\sc iii}], 
in very good agreement with each other. These strong gradients occur 
more or less at the same distance from the nucleus ($\sim$8\,\arcsec)
towards the north-east of the galaxy, almost at the position where the two
slits intersect (see Figs.\ \ref{hst-slits} and
\ref{hst-slits-3310-ACS}). The curves show a step in the radial velocity of
about 70 and 90\,km\,s$^{-1}$ for S1 and 
S2, respectively. In Fig.\ \ref{hst-slits-3310-ACS} we indicate with two 
circles of 8 and 18\arcsec\ radii the approximate position where this
step is located and the distance from the nucleus up to where we can
derive the radial velocities for S1. The stellar radial velocity of R1+R2
derived using the 5\,px aperture (criss-crossed white-blue diamond in the
lower panel of Fig.\ \ref{velocities-3310}) is in very good agreement
with the values derived from the gas emission lines. 

A very similar result was found by
\cite{2007A&A...469..405P} using high spatial resolution spectra
($\sim$\,0.07\,\arcsec\,px$^{-1}$) and moderate spectral resolution
(R\,=\,$\Delta\lambda$/$\lambda$\,$\sim$\,6000, yielding a resolution of
$\Delta\lambda$\,$\sim$\,1.108\,\AA\,px$^{-1}$ at H$\alpha$) from the Space
Telescope Imaging Spectrograph (STIS) on board the HST, and using a 
2\,$\times$\,2 on-chip binning for the non nuclear spectra. The position
angle of their slits is 170$^\circ$ and three parallel positions were used,
one of them passing across the nucleus and the other two with 0.2\,\arcsec\
offset to both sides \footnote{It must be noted that the 
H$\alpha$ narrow band-filter image in Fig.\ 2 of that work (the same ACS
image presented by us) is rotated by 180$^\circ$ with respect to their quoted
position, since the Jumbo region appears placed at the north-east of the
nucleus (according to the orientation of the image given by these authors)
while this region is located at the south-west 
\citep[see e.g.\
][]{1984ApJ...284..557T,1990MNRAS.242P..48T,1993MNRAS.260..177P}.}. 
As pointed out by Pastorini and collaborators, NGC\,3310 shows a typical
rotation curve expected for a rotating disc 
\citep[a typical $S$ feature,][]{2006A&A...448..921M} but the curve is
disturbed. Besides, we can appreciate that  
near the Jumbo region there is a deviation from circular rotation
motion. The step in the radial velocity found by us is an effect of the
spatial resolution. In Fig.\ 3 of \cite{2007A&A...469..405P}, the better
spatial resolution of STIS-HST resolves the steep gradient and
completes the information missing in our data. This disturbed behaviour of the
rotation curve 
of the gas in the galactic disc, characterized by non-circular motions, was
also found by \cite{1995A&A...300..687M} using H{\sc i} radio data;
\cite{1996A&A...309..403M} from H{\sc i} and H{\sc ii} radio
data and \cite{2001A&A...376...59K} using optical data. They found a
strong streaming along the spiral arms, supporting the 
hypothesis of a recent collision with a dwarf galaxy that triggered the
circumnuclear star formation during the last 10$^8$\,yr.
Besides, \cite{1976A&A....49..161V} shows that the rotation centre of the
gas is offset with respect to the stellar continuum (by $\sim$\,1.5\,\arcsec). 
The central velocity of NGC\,3310 derived by us is in very good agreement with
that previously estimated by
\cite{1976A&A....49..161V,1995A&A...300..687M,1996A&A...309..403M,1998AJ....115...62H,2001A&A...376...59K}.
The velocity distribution is also in very good agreement with that expected
for this type of galaxies \citep{1987gady.book.....B}. 


\begin{figure}
\centering
\includegraphics[width=.46\textwidth,angle=0]{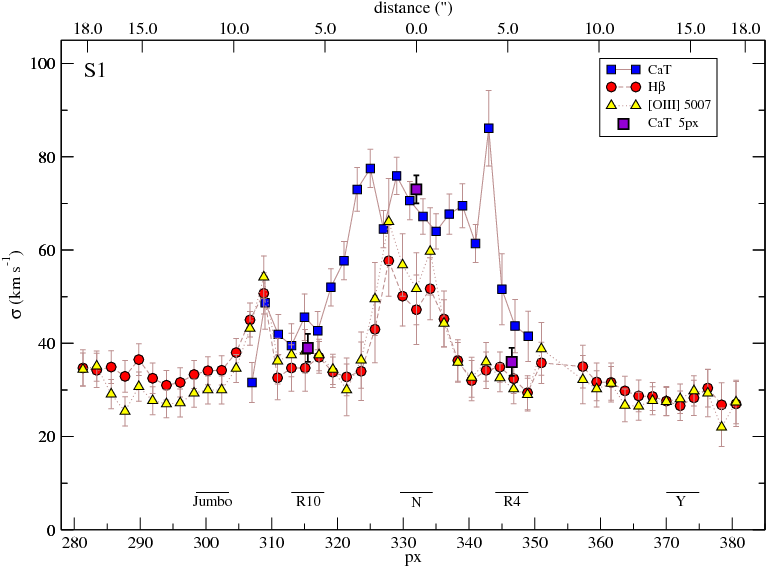}\\
\vspace*{0.3cm}
\includegraphics[width=.46\textwidth,angle=0]{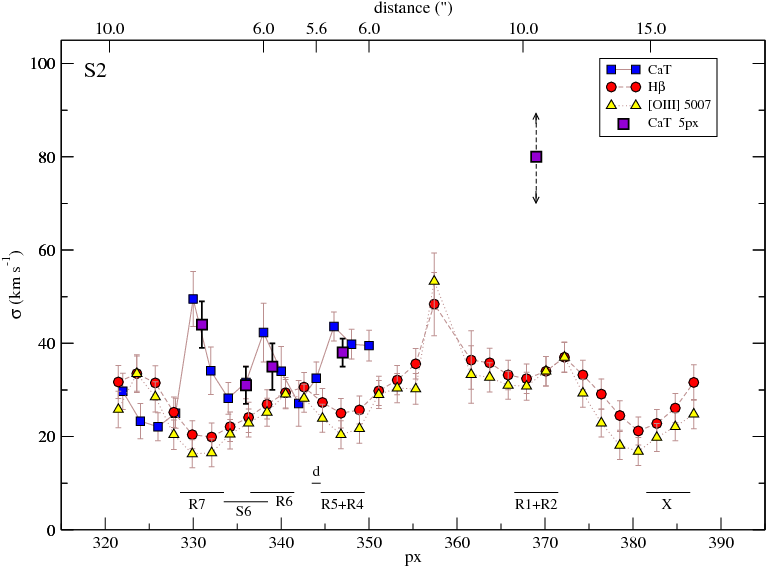}
\caption[]{Velocity dispersions along the slit versus pixel
  number for each slit position of NGC\,3310 (upper panel: S1; lower panel:
  S2) as derived from the gas emission lines (small red circles: H$\beta$;
  small yellow triangles: [O{\sc iii}]) and the stellar absorption ones (small
  blue squares). The stellar velocity dispersions derived for each region
  and the nucleus using the 5\,px aperture are also plotted with violet
  squares. The
  individual CNSFRs and the nucleus, ``N'', or the closest position to it,
  ``d'', are marked in the plots. The distance in arcsec from the nucleus is
  displayed in the upper x-axis of each panel. [{\it
  See the electronic edition of the Journal for a   
  colour version of this figure.}]}
\label{dispersions-px-3310}
\end{figure}

Fig.\ \ref{dispersions-px-3310} shows the run of velocity dispersion along the
slit versus pixel number for slit positions S1 and S2
respectively. These velocity dispersions have been derived from the gaseous
emission lines, H$\beta$ and [O{\sc iii}], and the stellar absorption ones
using 2 and 3\,px apertures for S1 and S2, respectively.
We have marked the location of the studied CNSFR and the galaxy
nucleus. 
We have also plotted the stellar velocity dispersion derived for each
region and the nucleus using 5\,px apertures. The values
derived with wider apertures are approximately the average of the velocity
dispersions estimated using the narrower ones. Therefore, the apparent
increase of the 
velocity dispersions of the regions lying on relatively steep parts
of the rotation curve by spatial ``beam smearing'' of the rotational velocity
gradient due to the finite angular 
resolution of the spectra is not so critical for the studied CNSFRs. This is
due to the fact that the turnover points of the rotation curves and the
star-forming ring seem to be located at the same positions, as was pointed out
above. The velocity dispersions, both for
stars and gas, show a behaviour characteristic of a regular circular motion in
a rotating disc, a smooth plateau and a central peak. A similar result
was found by \cite{2007A&A...469..405P}.

\subsection{Star cluster masses}

The eight CNSFRs observed in NGC\,3310 show a
complex structure at the HST resolution (Fig.\ \ref{sizes}), with a good
number of subclusters with linear diameters between 2.2 and 6.2 pc.
For these individual clusters, except for R1+R2, the derived upper limits to
the masses are in the range 
between 1.8 and 7.1\,$\times$\,10$^6$\,M$_\odot$ (see table 
\ref{mass}), with fractional errors between $\sim$\,16 and $\sim$\,32 per
cent. The upper limits to the dynamical masses estimated for the whole CNSFRs
(``sum''), except for R1+R2, are between  
2.1\,$\times$\,10$^7$ and 1.4\,$\times$\,10$^8$\,M$_\odot$, with fractional
errors between $\sim$3 and $\sim$8 per cent. 
The stellar velocity dispersion of region R1+R2
has a large associated error, so we do not quote an explicit error for its
derived mass which we consider highly uncertain. The masses for the individual
clusters of this CNSFR, derived using this value of the velocity dispersion
are larger, between 
1.2 and 2.4\,$\times$\,10$^7$\,M$_\odot$. However, if we assume that the
relation between the gas and the stellar velocity dispersions follows a
behaviour similar to that shown by the rest of the CNSFRs of NGC\,3310 (see
Fig.\ 
\ref{dispersions3310}) these masses could be reduced by a factor of 2.3,
yielding individual masses between 5.2\,$\times$\,10$^6$ and
1.0\,$\times$\,10$^7$\,M$_\odot$. In the same way, the total dynamical mass of
R1+R2, 9.1\,$\times$\,10$^8$\,M$_\odot$, would be reduced to
4.0\,$\times$\,10$^8$\,M$_\odot$.
The upper limits to the dynamical mass derived for the nuclear region inside
the inner 14.2\,pc is 
5.3\,$\times$\,10$^7$\,M$_\odot$, with a fractional error of about 11 per
cent. This value is somewhat higher than that derived by
\cite{2007A&A...469..405P} under the assumption of the presence of a super
massive black hole (SMBH) in the centre of the galaxy. When they allow for the
disc inclination, this mass varies in the range 
5.0\,$\times$\,10$^6$\,-\,4.2\,$\times$\,10$^7$\,M$_\odot$.

The masses of the ionizing stellar clusters of the CNSFRs,
derived from their H$\alpha$ luminosities (under the assumption that the
regions are ionization bound and without taking into account any photon
absorption by dust), are between 8.7\,$\times$\,10$^5$
and 2.1\,$\times$\,10$^6$\,M$_\odot$ for the star-forming regions, and
3.5\,$\times$\,10$^6$\,M$_\odot$ for the nucleus (see Table
\ref{parameters}). In the case of the Jumbo region this mass vary between 1.3
and 
3.1\,$\times$\,10$^6$\,M$_\odot$ whether we take the H$\alpha$ luminosities
from \cite{1993MNRAS.260..177P} or \cite{2000MNRAS.311..120D}
respectively, probably because the estimated reddening constants in these two
works are different (see Table \ref{parameters}). These different values of
the luminosities can be due to the very complex
structure of the region (see Fig.\ 1 of both works) and to different
selection of the zone used to measure the H$\alpha$ fluxes. Besides, Pastoriza
and colleagues used spectroscopic data while D\'iaz and collaborators used
photometric images.
In column 9 of Table \ref{parameters} we show a comparison 
(in percentage) between the ionizing stellar masses of the circumnuclear
regions and their dynamical masses. These values are approximately between
1\,-\,7 per cent for the CNSFRs, and 5 per cent for the nucleus of
NGC\,3310. 

Finally, the masses of the ionized gas, also derived from their
H$\alpha$ luminosities, range between 1.5 and
7.2\,$\times$\,10$^5$\,M$_\odot$ for the CNSFRs, and
4.7\,$\times$\,10$^4$\,M$_\odot$ for the nucleus (see Table 
\ref{parameters}). They make up a small fraction of the total mass of the
regions. 
As in the case of the masses of the ionizing stellar
clusters of the Jumbo region and for the same reasons, the mass of the
ionized gas derived using the H$\alpha$ luminosities from
\cite{1993MNRAS.260..177P} or from \cite{2000MNRAS.311..120D} are 
different, 3.93 and 9.52\,$\times$\,10$^5$\,M$_\odot$, respectively.
It should be taken into account that we have derived
both the masses of the ionizing stellar clusters and of the ionized gas from
the H$\alpha$ luminosity of the CNSFRs assuming that they consist of one
single component. However, if we consider only the broad component whose
kinematics follows that of the stars in the regions, all derived quantities
would be smaller by a factor of 2. 

Comparing the colours of the small-scale clusters from HST data with models,
\cite{2002AJ....123.1381E} suggest that most of these clusters contain
masses from 10$^4$\,M$_\odot$ to several times 10$^4$\,M$_\odot$. For the
large-scale ``hot-spots'' they estimate masses ranging from 10$^4$ to several
times 10$^5$\,M$_\odot$. For their 108+109 region (almost equivalent to the
Jumbo region) they derived a mass of 6.3\,$\times$\,$10^5$\,M$_\odot$, similar
to the value derived by \cite{1993MNRAS.260..177P},
7\,$\times$\,$10^5$\,M$_\odot$. 
They estimated from the unusually large H and He{\sc ii} emission line
luminosities, that this region must contain 220 WN and 570 OB
stars. Regarding other regions, Elmegreen and colaborators derived a mass of
$10^6$\,M$_\odot$ for 
their region 78 using the NIR data and 7.7\,$\times$\,$10^6$\,M$_\odot$ from
their optical ones. Besides, they found 17 candidate SSCs, with masses in the
range  2\,$\times$\,$10^4$ to 4.5\,$\times$\,$10^5$\,M$_\odot$.
Some of them coincide with the CNSFRs studied by us, such as 48, 49 and
50, approximately coincident with R5 and R6. These SSCs are mostly in the
innermost southern spiral arm, with some in the northern one or outside the
southern 
arm of the ring \citep{2002AJ....123.1381E}. Their derived (J-H) colours
suggest two different populations of SSCs, a very young of few million years
and an older one in a range between 10 and 50 million years, with the younger
clusters located in the northern part of the ring.


It should be recalled that we have estimated the dynamical masses
through the virial theorem under the assumption  that the systems are
spherically symmetric, gravitationally bound and have isotropic velocity
distribution. We have used the stellar velocity dispersions derived from the
CaT absorptions features and the cluster sizes measured from the high spatial
resolution WFPC2-PC1 HST image. As mentioned at the end of section 4, 
the use of these wider size scale velocity dispersion
measurements to estimate the mass of each knot, can lead to an overestimate of the
mass of the individual clusters, and hence of each CNSFR (see Paper I).

However, as can be seen in the HST-NICMOS images (right-hand panel of Fig.\
\ref{hst-slits}),
the CNSFRs are clearly visible in the IR
and dominate the light inside the apertures observed \cite[see][for a detailed
analysis of the IR images]{2002AJ....123.1381E}. All the regions analyzed
show up very prominently in the near-IR and therefore we can assume that the
light at the CaT wavelength region is dominated by the stars in the clusters.
The IR CaT is very strong, in fact the strongest stellar feature, in
young clusters older than 4\,Myr \cite{1990MNRAS.242P..48T}.  Besides,
we detect a minimum in the velocity dispersion at the position of most of the
clusters, indicating that they are kinematically distinct. We can not be sure
though that we are actually  
measuring their velocity dispersion and thus prefer to say that our
measurements of $\sigma_{\ast}$ and hence the dynamical masses constitute upper
limits. Although we are well aware of the difficulties, still we are confident
that these upper limits are valid and important for comparison with the gas
kinematic measurements.

Another important fact that can affect the estimated dynamical masses is the
presence of binaries among the red supergiant and red giant populations from 
which we have derived the stellar velocity dispersions. 
In a recent work, \cite{2009AJ....137.3437B} using
the Gemini Multi-Object Spectrograph (GEMINI-GMOS) data, have investigated the
presence of binary  
stars within the ionizing cluster of 30~Doradus. From a seven epoch observing
campaign they have detected a rate of binary system candidates within their OB
star sample of $\sim$\,50\,\%. Interestingly enough,
this detection rate is consistent with a spectroscopic population of 100\,\%
binaries, when the observational parameters described in the simulations by
\cite{2001RMxAC..11...29B} are set for their observations. 
From their final sample of `single' stars 
they estimated a radial velocity dispersion of 8.3\,km\,s$^{-1}$. When they
derived $\sigma_{\ast}$ from a single 
epoch, they found values as high as 30\,km\,s$^{-1}$,
consistent with the values derived from  single epoch NTT observations by
\cite{2001A&A...380..137B}. 

Although the environment of our CNSFRs is very different from that of 30~Dor
and the stellar components of the binary systems studied by Bosch et al. (2009)
are very different from  the stars present 
in our regions from where the CaT arise (red supergiants), this is an
illustrative observational example of the problem. The orbital motions of
the stars in binary (multiple) systems produce an overestimate of the
velocity dispersions and hence of the dynamical masses. The single-star
assumption introduces a systematic error that depends on the properties of the
star cluster and the binary population, with an important effect on the
cluster mass if the typical orbital velocity of a binary component is of
the order of, or larger than, the velocity dispersion of the single/binary
stars in the potential of the cluster \citep{2008A&A...480..103K}. As was
pointed out by these authors, the
relative weights between the single and binary stars in the velocity dispersion
measurements depend on the binary fraction, which, together with the
semi-major axis or period distribution, are the most important parameters in
order to 
determine if the binary population affects  the estimated dynamical
masses. Their simulations indicate that the dynamical mass is overestimated by
70\,\%, 50\,\% and 5\,\% for a measured stellar velocity dispersion in the
line of sight of 1\,km\,s$^{-1}$, 2\,km\,s$^{-1}$ and 10\,km\,s$^{-1}$
respectively. They therefore conclude that most of the known dynamical masses 
of massive star clusters are only mildly affected by the presence of
binaries. Hence, although the binary
fraction of the red supergiants and red giants in our circumnuclear
regions is unknown, since
the smallest estimated velocity  
dispersion is 31\,km\,s$^{-1}$, we can assume that the contribution of
binaries to the stellar velocity dispersions is not important.

\section{Summary and conclusions}

We have measured gas and stellar velocity dispersions in eight CNSFRs and
the nucleus of the barred spiral galaxy NGC\,3310. The stellar velocity
dispersions have been measured from the CaT lines at
$\lambda\lambda$\,8494, 8542, 8662\,\AA, while the gas velocity dispersions
have been measured by Gaussian fits to the H$\beta$\,$\lambda$\,4861\,\AA\ and
the [O{\sc iii}]\,$\lambda$\,5007\AA\ emission lines on high dispersion
spectra.

Stellar velocity dispersions are between 31 and 73\,km\,s$^{-1}$. The stellar 
and gas velocity dispersion are in relatively good agreement, with the former
being 
slightly larger. The velocity dispersions from [O{\sc iii}]\,5007\,\AA\ behave
similarly to those from H$\beta$.
However, the best Gaussian fits involve two different components for the gas:
a ``broad component" with a velocity dispersion larger than that measured for
the stars by about 20\,km\,s$^{-1}$, and a ``narrow component" with a velocity
dispersion lower than the stellar one by about 30\,km\,s$^{-1}$. This last
velocity component shows a relatively constant value for the two gas emission
lines, close to 22\,km\,s$^{-1}$ and also close to that measured for CNSFRs in
the previously studied NGC\,3351 and NGC\,2903. {\bf The velocity dispersion for the broad component of H$\beta$ is
similar to that of [OIII], contrary to what we found in Papers I and II for NGC~2903 and NGC~3351.}
We find a velocity shift between the narrow and broad
components of the multi-Gaussian fits that vary between -25 and
20\,km\,s$^{-1}$ in radial velocity.

When plotted in a [O{\sc iii}]/H$\beta$ versus [N{\sc ii}]/H$\alpha$ diagnostic
diagram, the broad and narrow  systems and those values derived using a single Gaussian
fit show very similar values, lying in the region of low-metallicity
\HII-like objects. This is in contrast to what we found in Papers~I and
II where the two systems are clearly segregated for the high-metallicity 
regions of NGC\,3351 and NGC\,2903, with the narrow component showing lower
excitation and being among the lowest excitation line ratios detected within
the SDSS data set of starburst systems.

The rotation curve of NGC\,3310 shows a typical $S$ feature, with
the presence of some perturbations, particularly near the location of the
Jumbo region. The values derived from the gas emission lines
and the stellar absorption features are in very good agreement. The
position going through the nucleus shows maximum and minimum values more or
less coincident with the location of the CNSFRs.

The upper limits to the dynamical masses estimated from the stellar velocity
dispersion using the virial theorem for the CNSFRs are in the
range between 2.1\,$\times\,10^7$ and 1.4\,$\times\,10^8$\,M$_\odot$. For
the nuclear region inside the inner 14.2\,pc this upper limit is
5.3\,$\times\,10^7$\,M$_\odot$. 
The upper limits
to the derived masses for the individual clusters are between 1.8 and
7.1\,$\times$\,10$^6$\,M$_\odot$.

Masses of the ionizing stellar clusters of the CNSFRs have been derived from
their H$\alpha$ luminosities under the assumption that the regions are
ionization bound and without taking into account any photon absorption by
dust. These masses derived for the star forming complexes of NGC\,3310 are
between 8.7\,$\times\,10^5$ and 2.1\,$\times\,10^6$\,M$_\odot$, and is 
3.5\,$\times\,10^6$ for the nucleus (see table \ref{parameters}). Therefore,
the ratio of the ionizing stellar population to the total dynamical mass is
between 0.01 and 0.07.

Derived masses for the ionized gas, also from their H$\alpha$ luminosities,
vary between 1.5 and 7.2\,$\times\,10^5$\,M$_\odot$ for the CNSFRs. For the
nucleus, the derived mass of ionized gas is 4.7\,$\times\,10^4$\,M$_\odot$.

It is interesting to note that, according to our findings, the SSC in CNSFRs
seem to contain composite stellar populations. Although the youngest one
dominates the UV light and is responsible for the gas ionization, it
constitutes only about 10 per cent of the total mass. This can explain the low
EWs of emission lines measured in these regions.  This may well apply to
studies of other SSC and therefore conclusions drawn from fits of single
stellar population models should be taken with caution 
\citep[e.g.\ ][]{2003ApJ...596..240M,2004AJ....128.2295L}.
Furthermore, the composite
nature of the CNSFRs  means that star formation in the rings is a process that
has taken place over time periods much longer than those implied by the
properties of the ionized gas.

\section*{Acknowledgements}
\label{Acknoledgement}
We acknowledge fruitful discussions with Guillermo Bosch, Nate Bastian and
Almudena Alonso-Herrero.
We are also grateful to an anonymous referee for his/her constructive
comments and revision of our manuscript. 

The WHT is operated in the island of La Palma by the Isaac Newton Group
in the Spanish Observatorio del Roque de los Muchachos of the Instituto
de Astrof\'\i sica de Canarias. We thank the Spanish allocation committee
(CAT) for awarding observing time.

Some of the data presented in this paper were obtained from the
Multimission Archive at the Space Telescope Science Institute (MAST). STScI is
operated by the Association of Universities for Research in Astronomy, Inc.,
under NASA contract NAS5-26555. Support for MAST for non-HST data is provided
by the NASA Office of Space Science via grant NAG5-7584 and by other grants
and contracts.

This research has made use of the NASA/IPAC Extragalactic Database (NED) which
is operated by the Jet Propulsion Laboratory, California Institute of
Technology, under contract with the National Aeronautics and Space
Administration and of the SIMBAD database, operated at CDS,
Strasbourg, France. 

This work has been supported by DGICYT grants AYA-2004-02860-C03 and
AYA2007-67965-C03-03. GH and MC
acknowledge support from the Spanish MEC through FPU grants AP2003-1821 and
AP2004-0977. Partial support from the Comunidad
de Madrid under grant S-0505/ESP/000237 (ASTROCAM) is acknowledged. Support
from the Mexican Research Council (CONACYT) through grant 19847-F is
acknowledged by ET and RT.

\bibliographystyle{mn2e}
\bibliography{ngc3310}

\end{document}